\documentclass[aps,prb,twocolumn,groupedaddress]{revtex4-1}

\usepackage{graphicx}
\usepackage{dcolumn}
\usepackage{bm}

\begin{document}

\title{Hydrogenated graphene on the silicon dioxide surface}

\author{P. Havu}
\email{Paula.Havu@aalto.fi}
\author{M. Ij\"{a}s}
\author{A. Harju}
\affiliation{Department of Applied Physics and Helsinki Institute of Physics, Aalto University School of Science, Espoo, Finland}

\begin{abstract}

Hydrogenation of graphene on the $\alpha$-quartz  (0001) SiO$_2$ substrate is studied, considering different surface terminations in order to take into account the amorphic nature of the material. Our \emph{ab initio} calculations show that the formation of graphane by hydrogen adsorption on graphene is energetically favored on hydroxyl and oxygen terminated  surfaces, whereas silicon termination and reconstruction of the oxygen termination hinder adsorption. Our results indicate that in order to fabricate graphane on SiO$_2$, it is beneficial to oxygenize the surface and saturate it with hydrogen. For the pristine graphene on the substrate, we find only marginal changes in the low-energy band structure for all surface terminations. 
\end{abstract}

\maketitle

\section{Introduction}

The two-dimensional allotrope of carbon, graphene, has many unique properties. For example, the dispersion is linear at the Fermi level, which makes the material semimetallic. For some applications like field-effect transistors, a band gap would be more desirable. Luckily, there are ways to induce a gap, like cutting graphene into narrow nanoribbons.\cite{Han}  An alternative approach is {the} chemical functionalization. In particular, hydrogen adsorption is {an} interesting {option} for two reasons: firstly, a band gap is induced \cite{Balog, Sofo} and secondly, the hydrogenated graphene sheet could serve as {a} hydrogen storage.\cite{Boukhvalov}

In experiments, the graphene layer is typically placed on a substrate, such as silicon dioxide,\cite{Elias, Ryu} silicon carbide   \cite{Sessi, Guisinger, Balog2} or a metal surface.\cite{Balog,Haberer}  The layer is then exposed to hydrogen by using hydrogen plasma \cite{Elias} or atomic hydrogen.\cite{Ryu, Balog, Balog2, Sessi, Haberer, Guisinger} The dense structure of the carbon layer suggests that hydrogenation should occur only on one side of the carbon plane.\cite{Guisinger} The hydrogen coverage reached in experiments varies and is difficult to estimate. Sessi \emph{et al.}~\cite{Sessi} demonstrated a quite dense and even hydrogen adsorption on SiC. Guisinger~\emph{et al.}~\cite{Guisinger} reached only a ``very low'' hydrogen coverage and Balog~\emph{et al.}~ \cite{Balog2} were not able to estimate the coverage on the same substrate. On SiO$_2$, neither Elias~\emph{et al.}~\cite{Elias} nor Ryu~\emph{et al.}\cite{Ryu} report the coverage. On the metal substrates, Haberer~\emph{et al.}~\cite{Haberer} find a maximum hydrogen coverage of 25\% and Balog~\emph{et al.}~\cite{Balog} report 3--27\%. The band gap of substrate-deposited graphane has not been accurately measured but a lower limit of approximately 0.5~eV has been estimated \cite{Balog, Elias}. Interestingly, in addition to hydrogenation, Sessi~\emph{et al.}~\cite{Sessi} also demonstrated selective dehydrogenation of the structure that restores the properties of pristine graphene. 

This far, theoretical studies on hydrogenated graphene, have concentrated on the suspended layer.\cite{Sofo, Samarakoon-Wang, Zhou-Wang, Zhou-Wu, Flores, Bhattacharya, Sahin-Ataca-Ciraci, Chandrachud, Artyukhov, Leenaerts,Lebegue} Graphane, with a hydrogen atom attached to every carbon atom, is found to be a good insulator with a large energy gap of even 6 eV in theoretical calculations.\cite{Sahin, Lebegue} Instead, a freestanding graphene membrane hydrogenated only on one side (called also graphone \cite{Zhou-Wang}), has not been found stable.\cite{Zhou-Wang, Zhou-Wu}  This opens the question of the role of the substrate for the structural and electronic properties of graphane.

In theoretical calculations, the presence of a substrate is widely neglected even in the case of pristine graphene.  Additionally, the theoretical descriptions\cite{Shemella-Nayak, Cuong-Otani-Okada, Kang, Hossain, Ao} of graphene on the SiO$_2$ substrate are contradictory even as to whether or not graphene forms bonds to the substrate, destroying the linear dispersion at the corners of the Brillouin zone.  To the best of our knowledge, the hydrogenation of graphene on SiO$_2$ has not yet been addressed.

In this work, we study the hydrogenation of graphene on SiO$_2$ using \emph{ab initio} calculations. We consider several different surface terminations in order to yield insight into the hydrogenation process of the real amorphic SiO$_2$.  Besides the simply cleaved oxygen terminated (OT) and cleaved Si terminated (SiT) surfaces, we also consider the reconstructed oxygen terminated (ROT) and hydrogen saturated oxygen terminated (OHT) surfaces, since the OT surface is reactive and prone to reconstruction or hydrogen saturation in ambient conditions.\cite{Goumans, Rignanese, Rignanese-Charlier}  We consider hydrogen coverages of graphene ranging from 1/8 to 3/4 monolayers (ML). Our calculations show that the hydrogen atoms are most likely to attach onto graphene on the OHT and OT surfaces. On the OHT surface, the arrangement of the hydrogen atoms in the lowest-energy configuration resembles the chair configuration of freestanding graphane. On the OT surface, hydrogen atoms form lines and make structure dispersive mainly in one direction. On the ROT or SiT substrates, we do not find stable hydrogen configurations. Our results suggest that in order to facilitate hydrogen adsorption on graphene, the substrate should be oxygenated and hydrogen passivated.  

\section{Methods}

{Our} calculations were performed using density-functional theory (DFT) with a van der Waals (vdW) correction,\cite{Tkatchenko} implemented in the all-electron FHI-aims code.\cite{AIMS} Double numeric plus polarization basis set of numerical atom-centered orbital basis functions and PBE \cite{PBE} exchange correlation functionals were used.  The structural relaxation was converged until the forces acting on the atoms were less than 0.001~eV/\AA{} and the electronic degrees of freedom were converged below $10^{-6}$~eV. Due to the abundance of metastable intermediate states, tight convergence criterions were found to be important. In the relaxations, no spin polarization was taken into account. For some of the final geometries, however, we searched spin moments but did not find any spin polarization. Of course, this does not preclude the possibility of more complicated geometries with finite spin moments. 

Silicon dioxide in the $\alpha$ phase was modelled using three SiO$_2$ unit cell thick slabs, corresponding to a layer of 15.8~\AA{}. Periodic boundaries were applied and a vacuum layer of approximately 20~\AA{} was placed between the periodic images of the slabs.  Four different SiO$_2$ surfaces were created: SiT, OT, ROT, and OHT surfaces.  The starting coordinates for the relaxations for the ROT surface were taken from the work of Goumans et al.\cite{Goumans} The lowest-energy OHT surface was obtained by relaxing the OT surface in the presence of additional hydrogen atoms, using different initial positions for the surface hydrogens.

The graphene layer was then placed on both sides of the silicon dioxide slab, the two interfaces being equivalent but rotated by 60$^{\circ}$.  The lattice mismatch between the unit vectors of graphene and SiO$_2$ was only 1.3~\%. In each SiO$_2$ unit cell, there were four graphene unit cells corresponding to eight carbon atoms (2$\times$2 carbon supercell). A 6$\times$6 $k$-point grid was used for this supercell.  The carbon atoms and three uppermost substrate atom layers were relaxed.  Different initial positions for the carbon grid were used and they were all found to relax to the same state. 

\begin{figure*}

\begin{tabular}{lrr} 
{\includegraphics[width=0.35\textwidth]{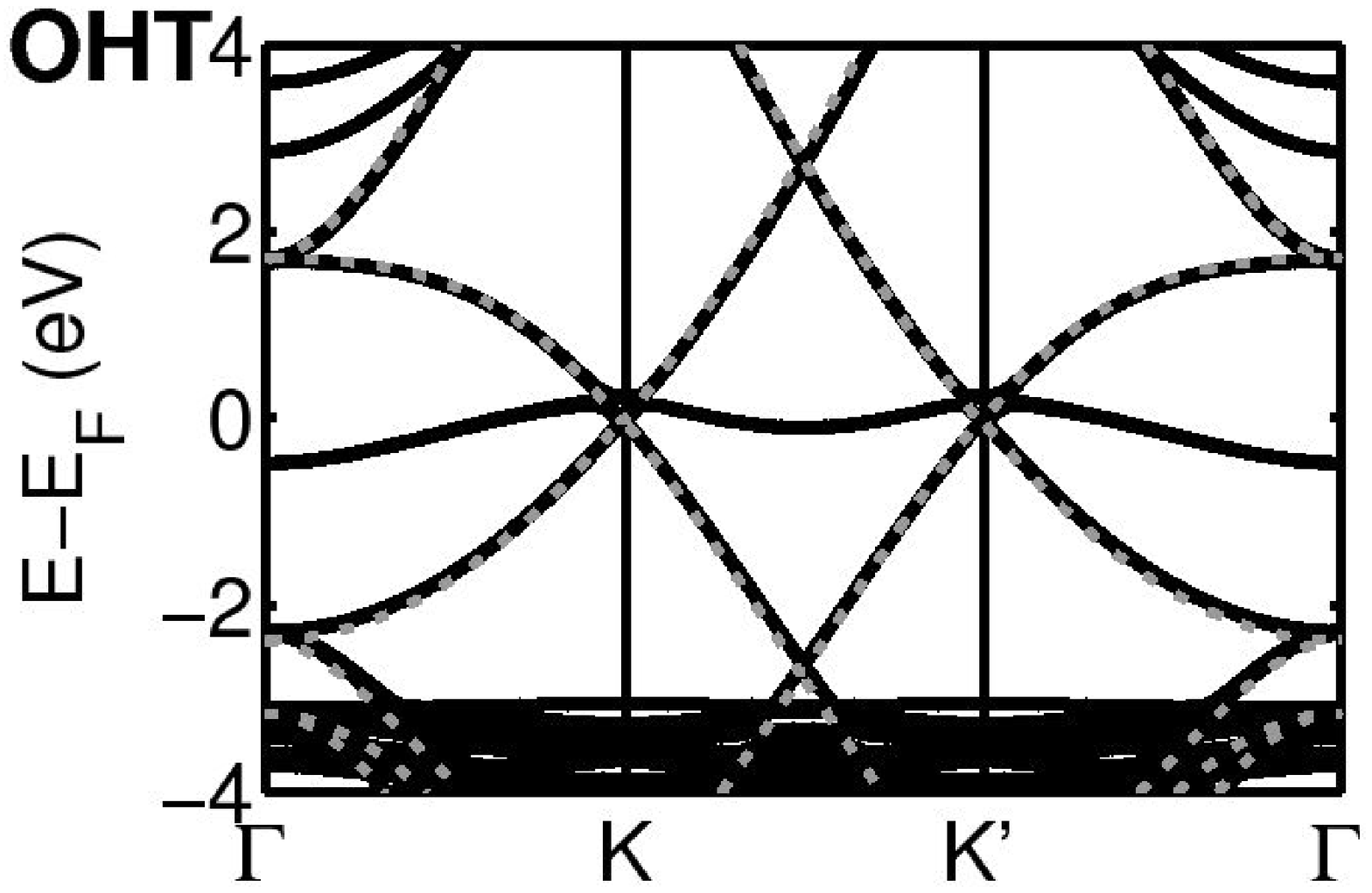}}& 
{\includegraphics[width=0.26\textwidth]{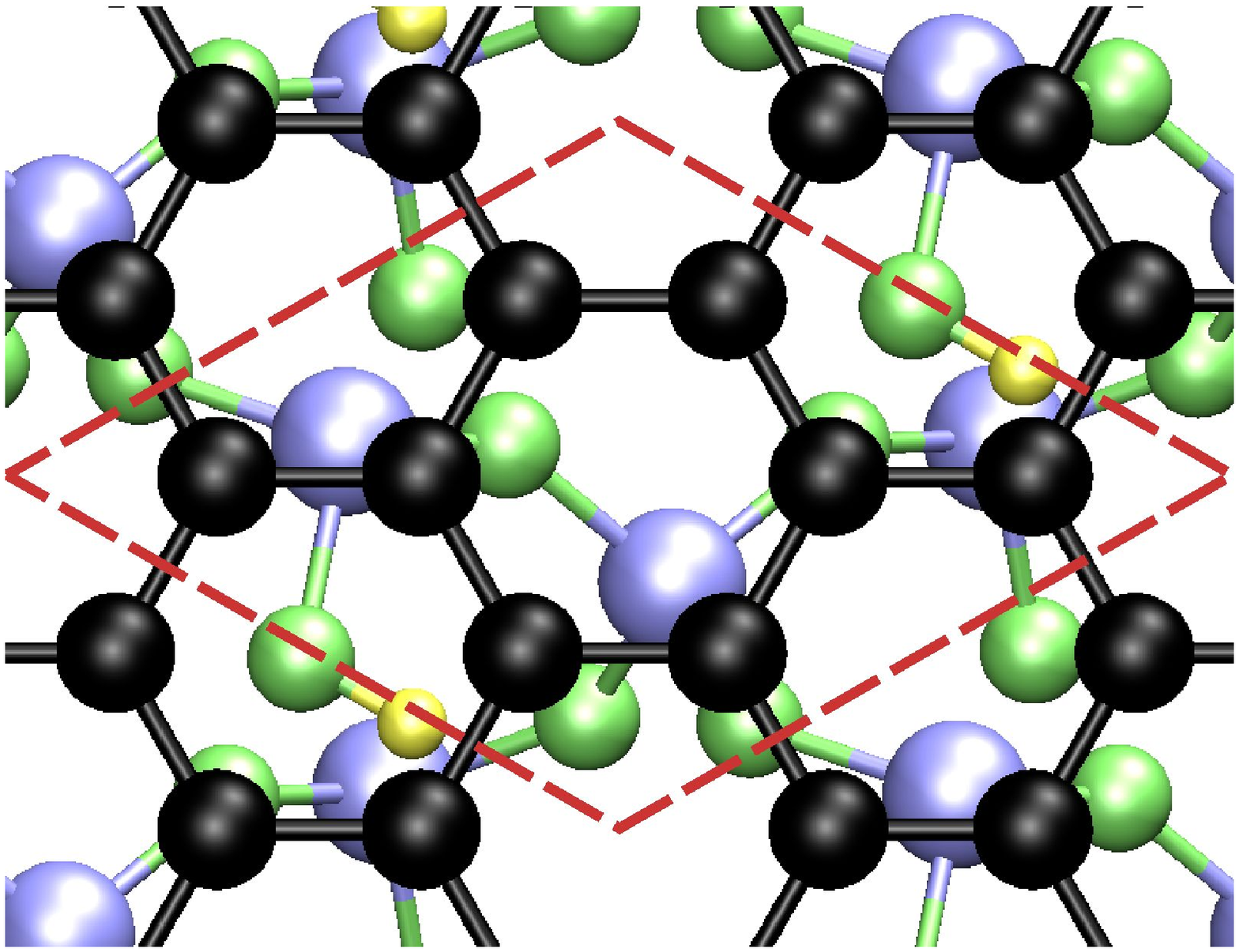}}&
{\includegraphics[width=0.29\textwidth]{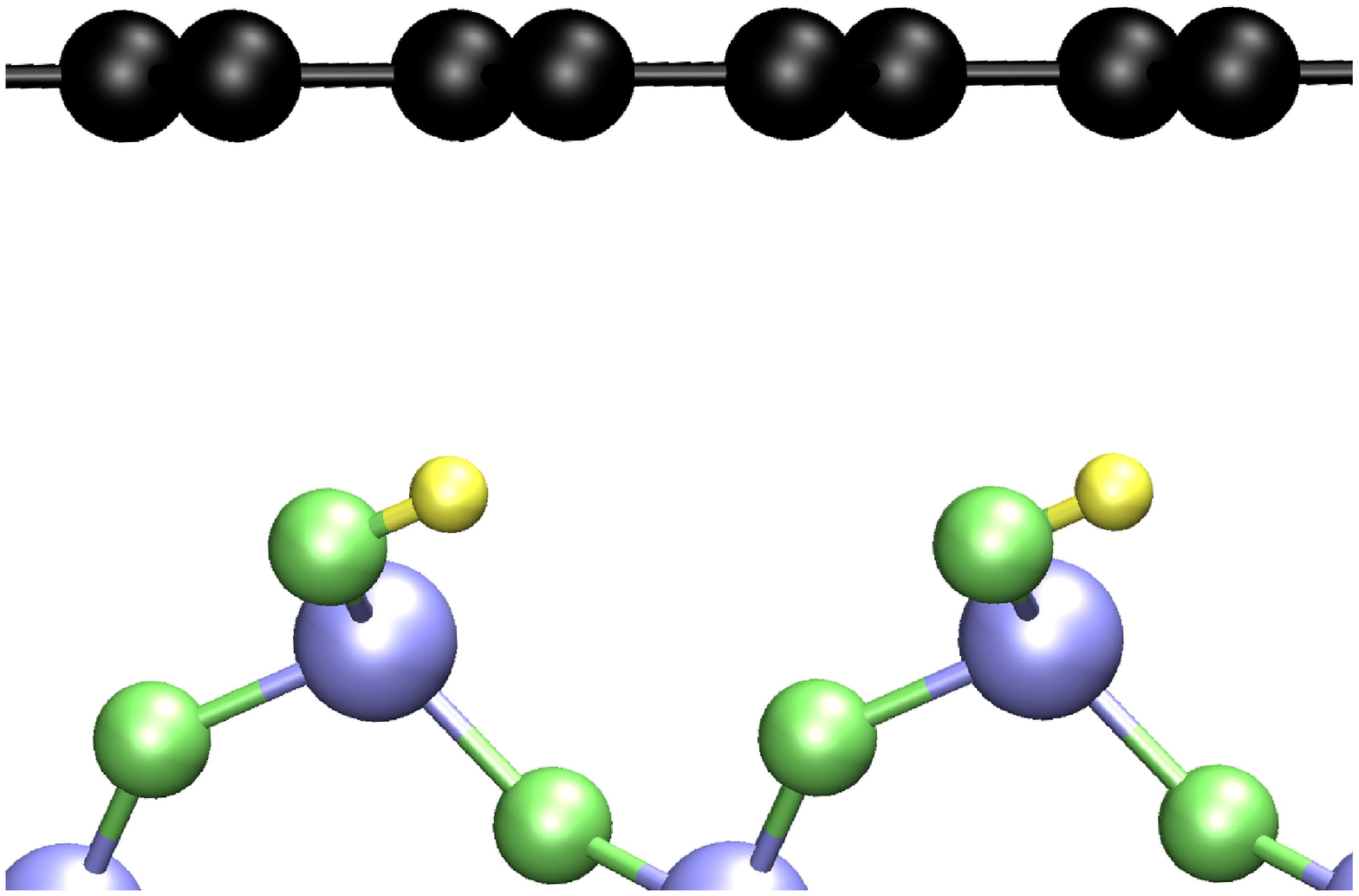}} \\
 {\includegraphics[width=0.35\textwidth]{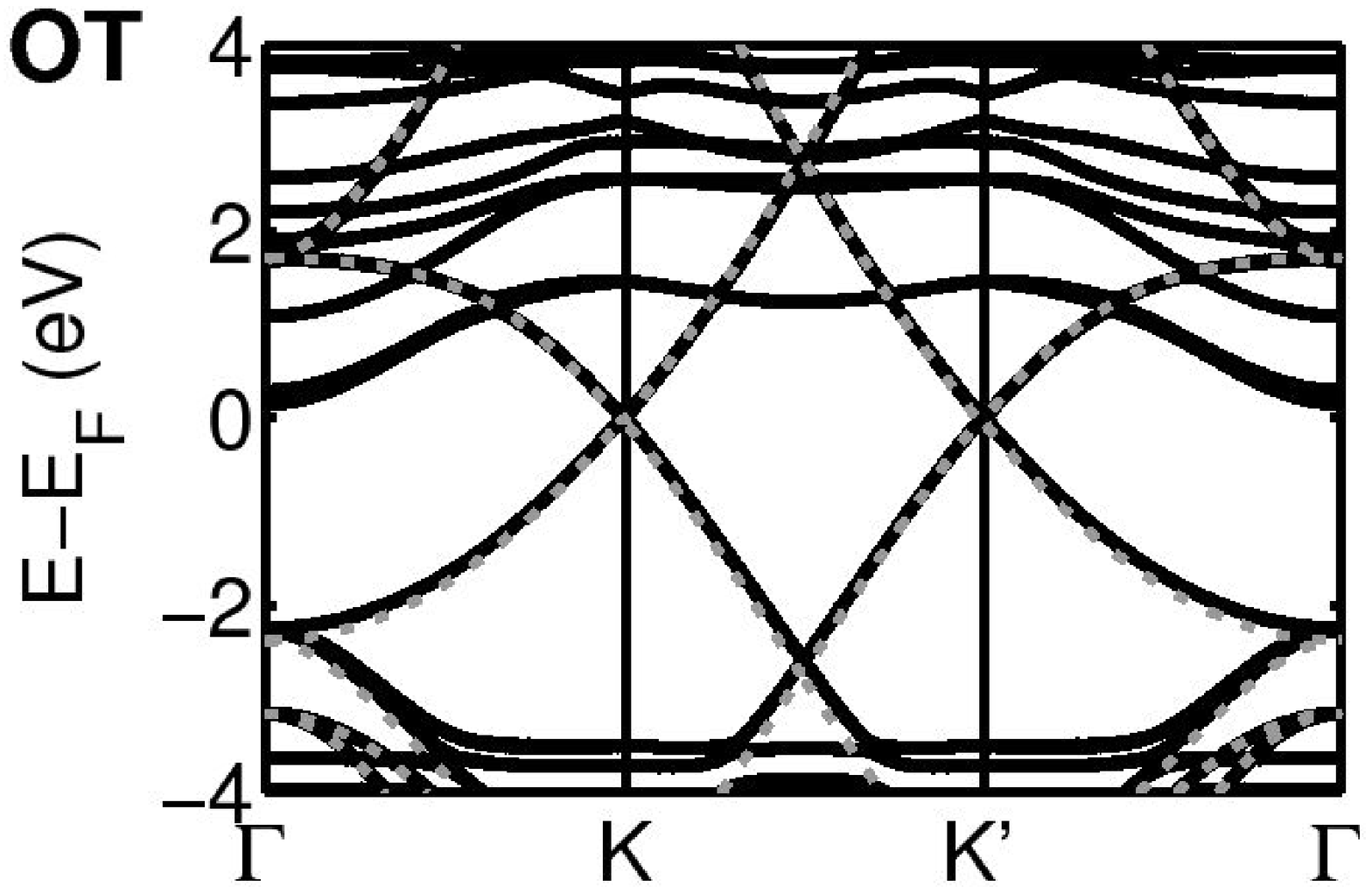}}&
{\includegraphics[width=0.26\textwidth]{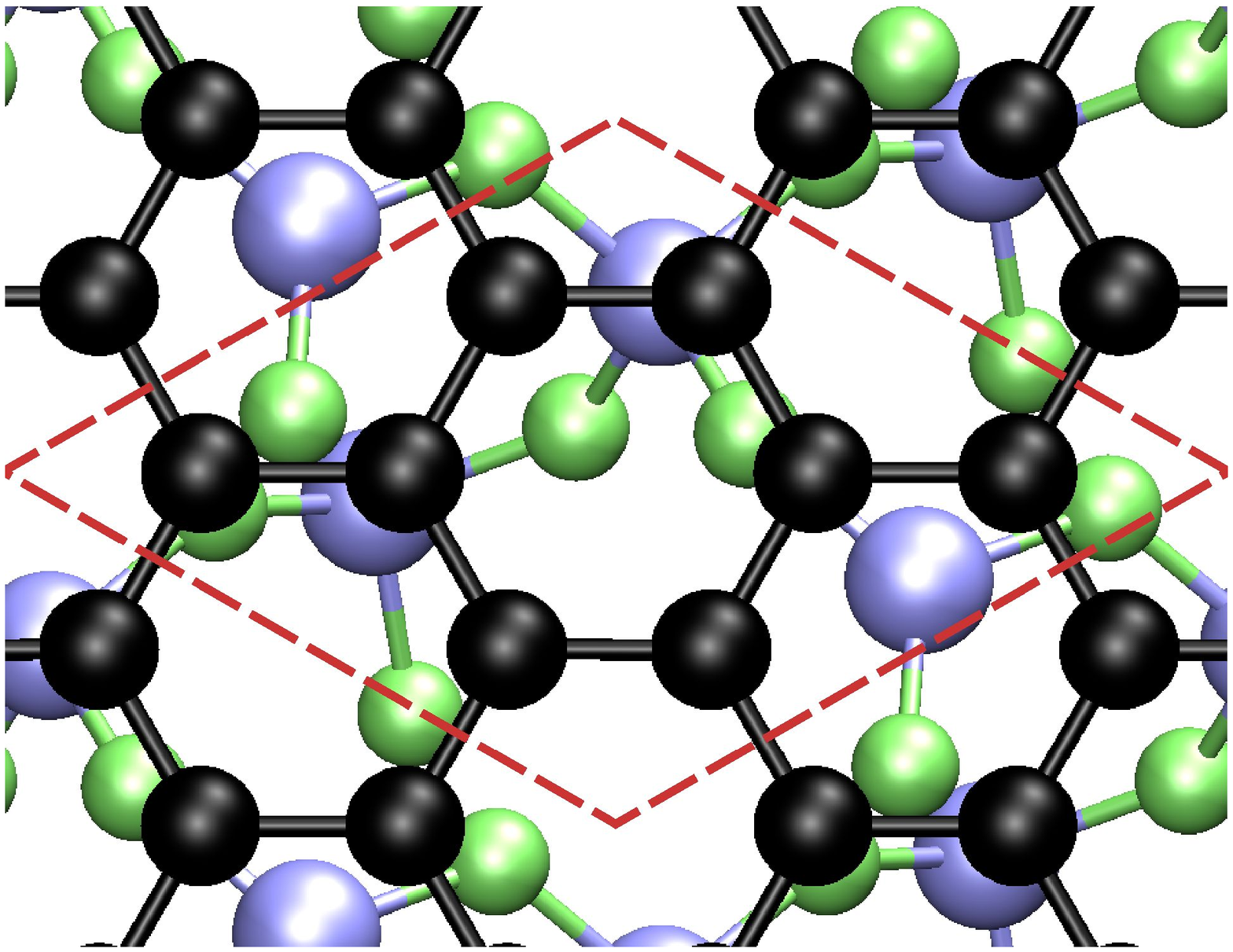}}&
{\includegraphics[width=0.29\textwidth]{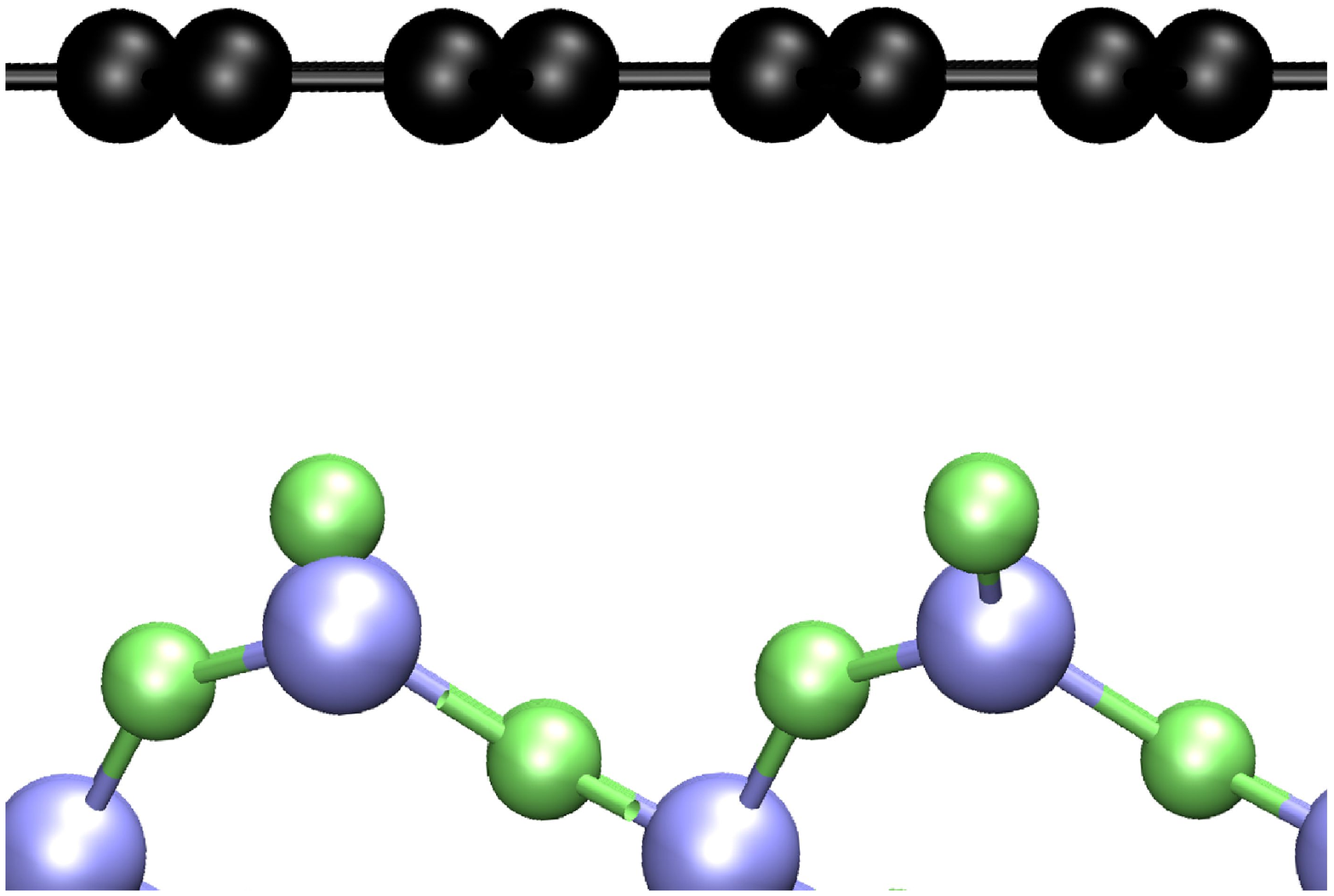}}\\
{\includegraphics[width=0.35\textwidth]{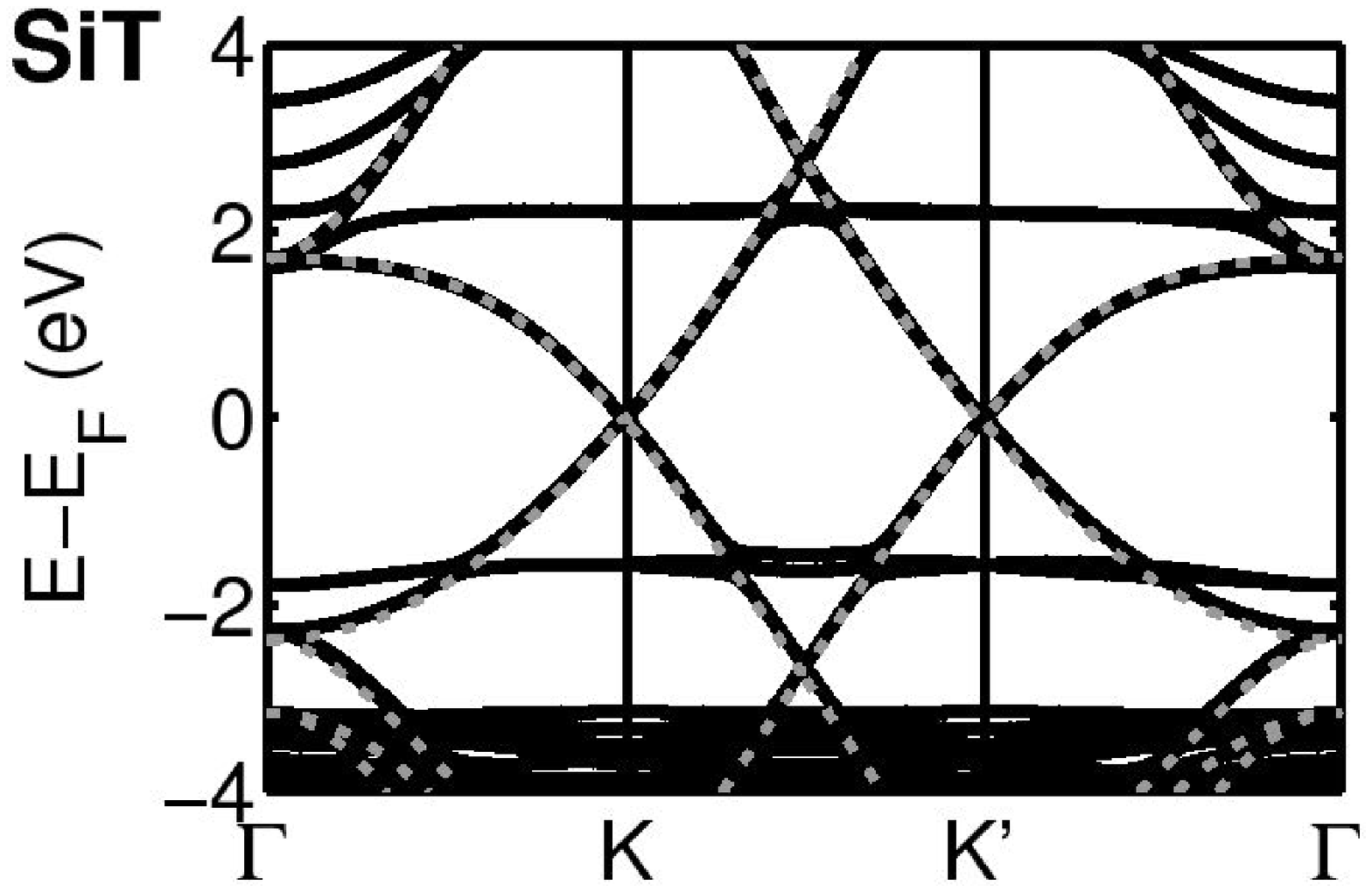}}& 
{\includegraphics[width=0.26\textwidth]{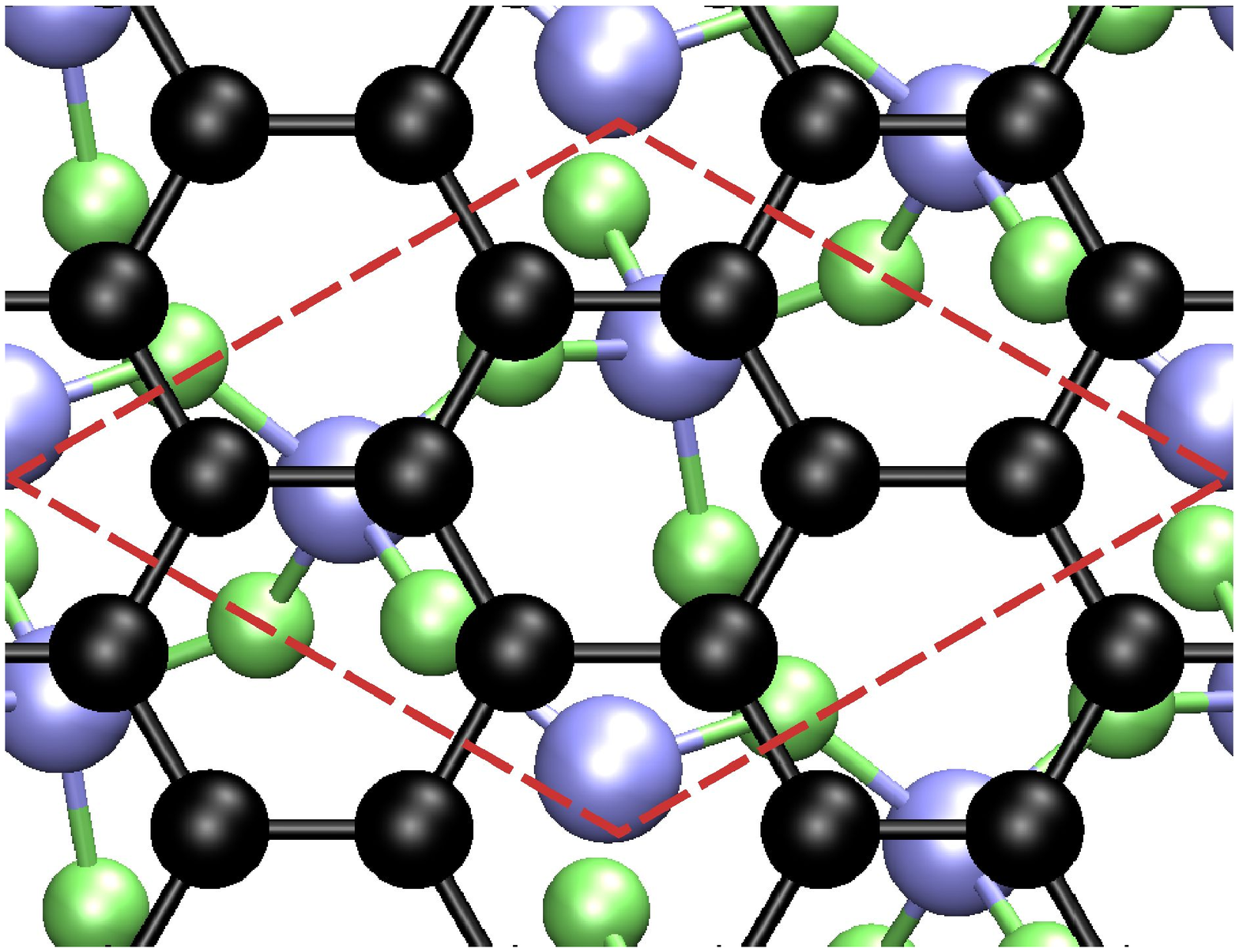}}&
{\includegraphics[width=0.29\textwidth]{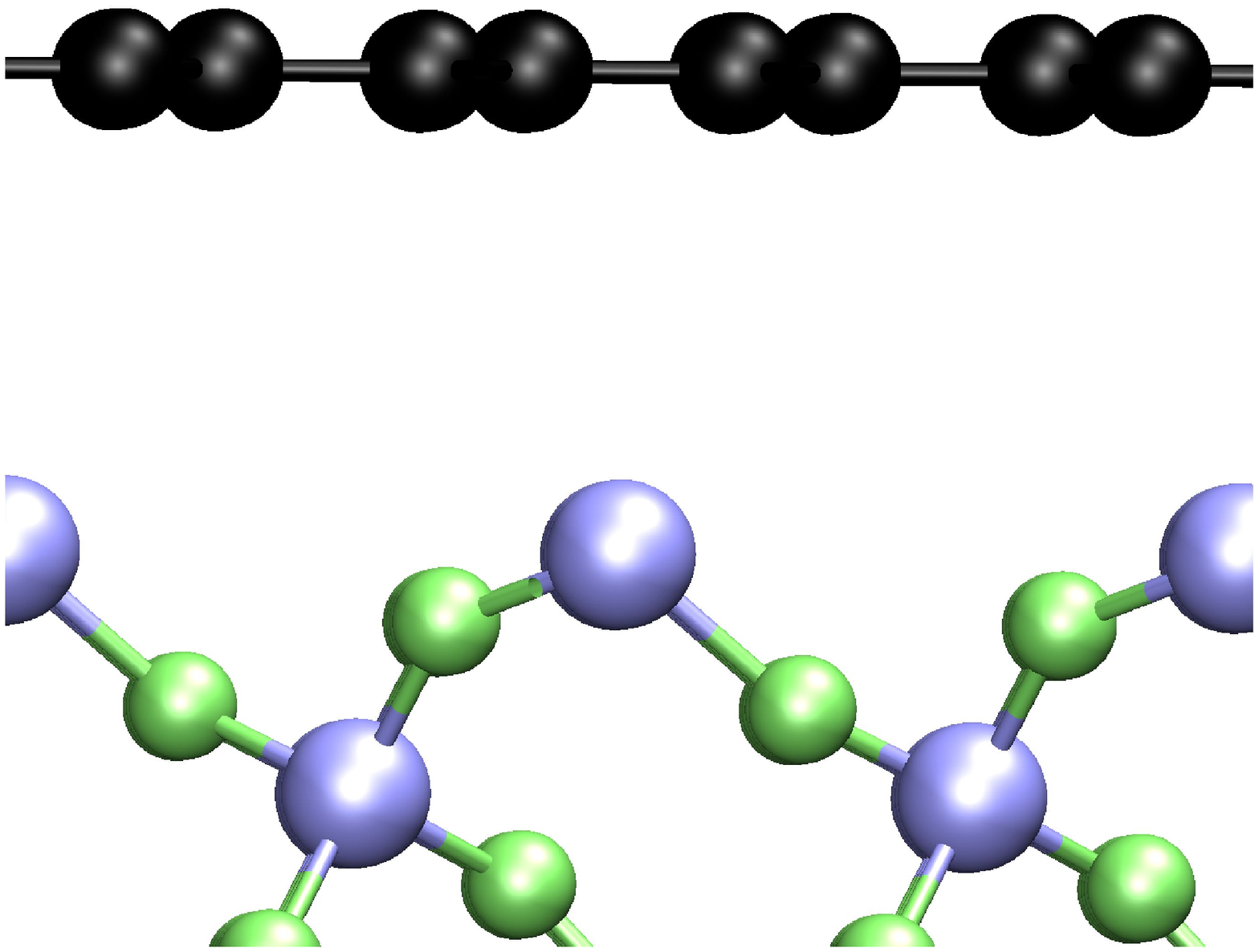}}\\
{\includegraphics[width=0.35\textwidth]{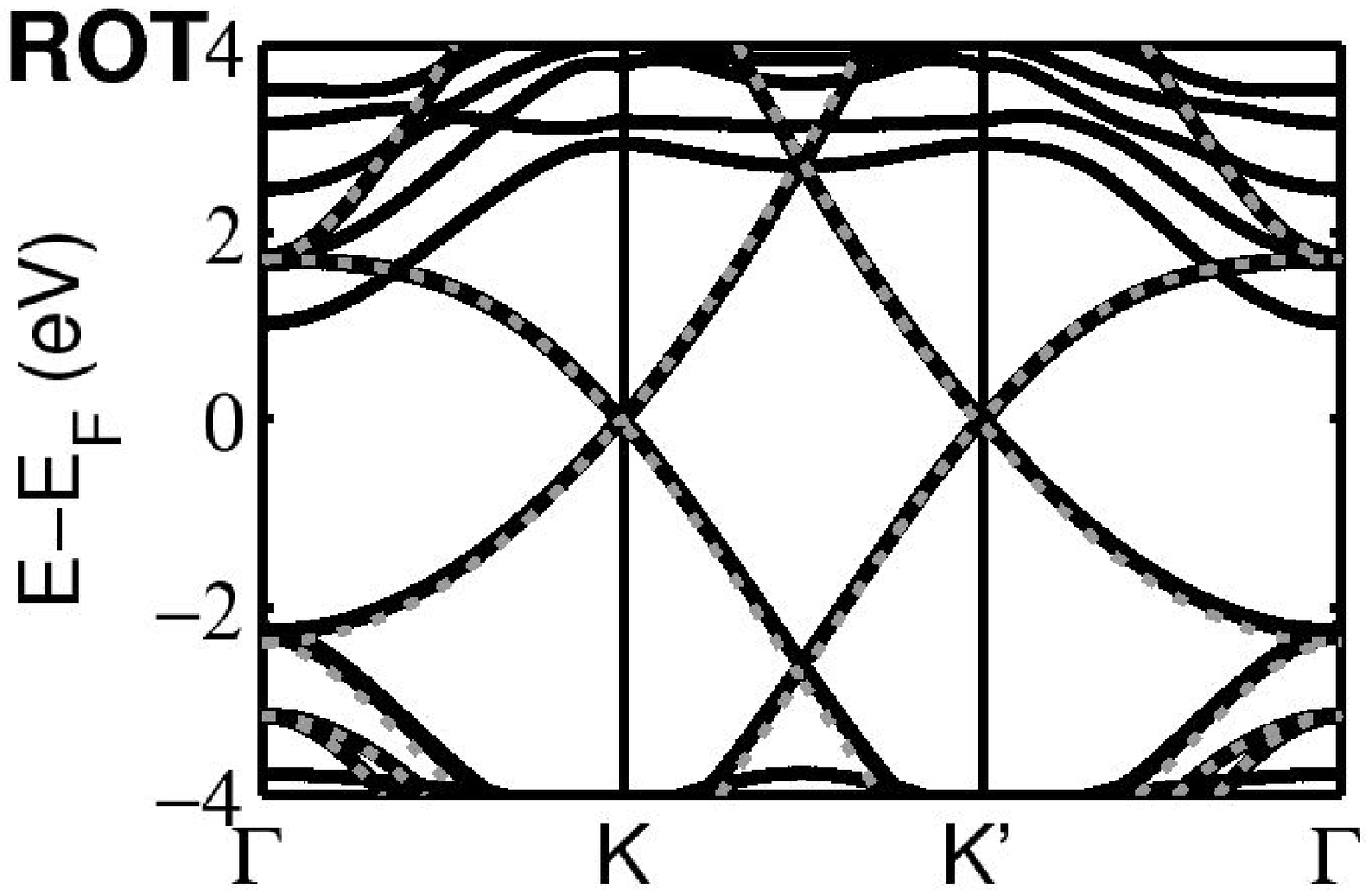}}&
{\includegraphics[width=0.26\textwidth]{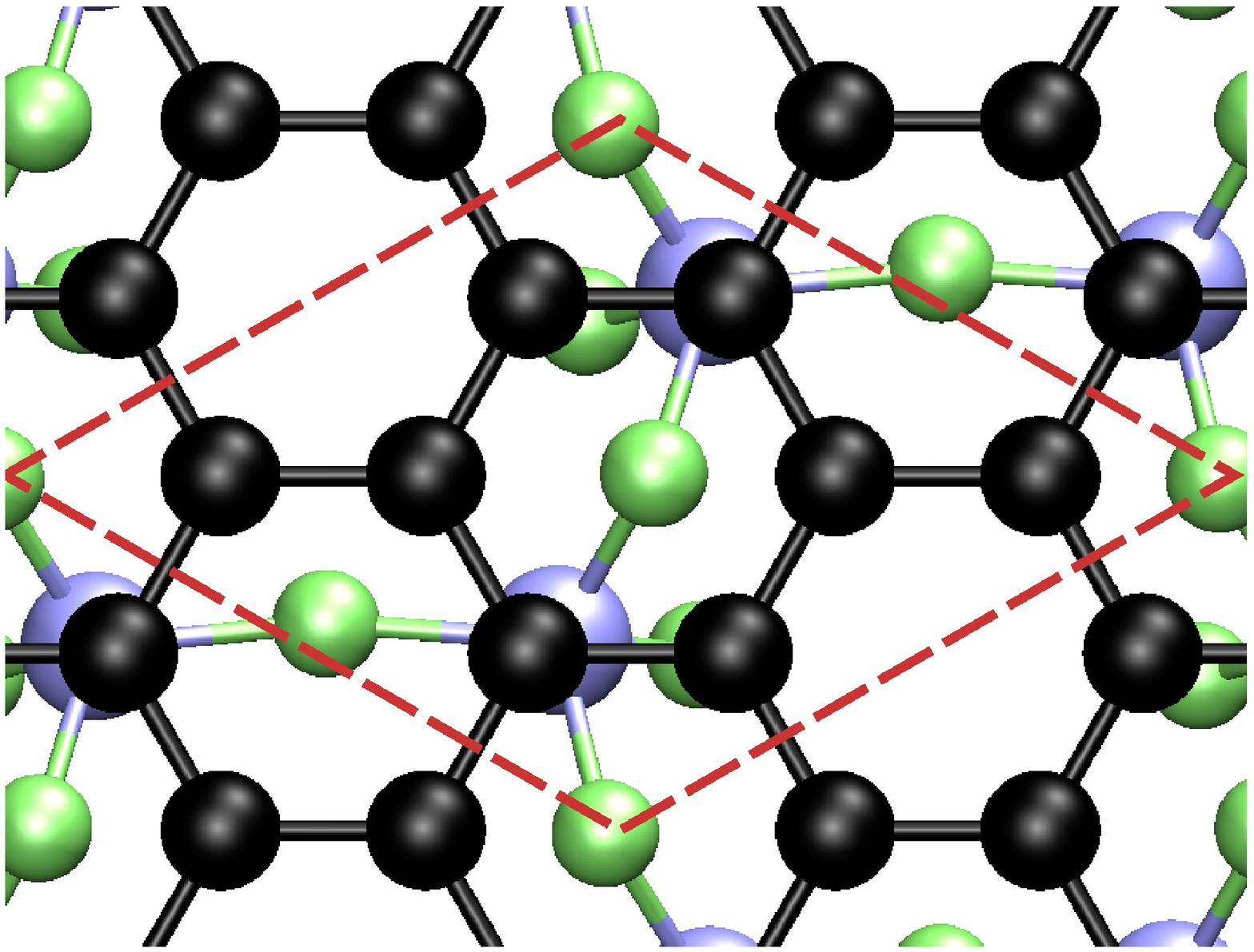}}&
{\includegraphics[width=0.29\textwidth]{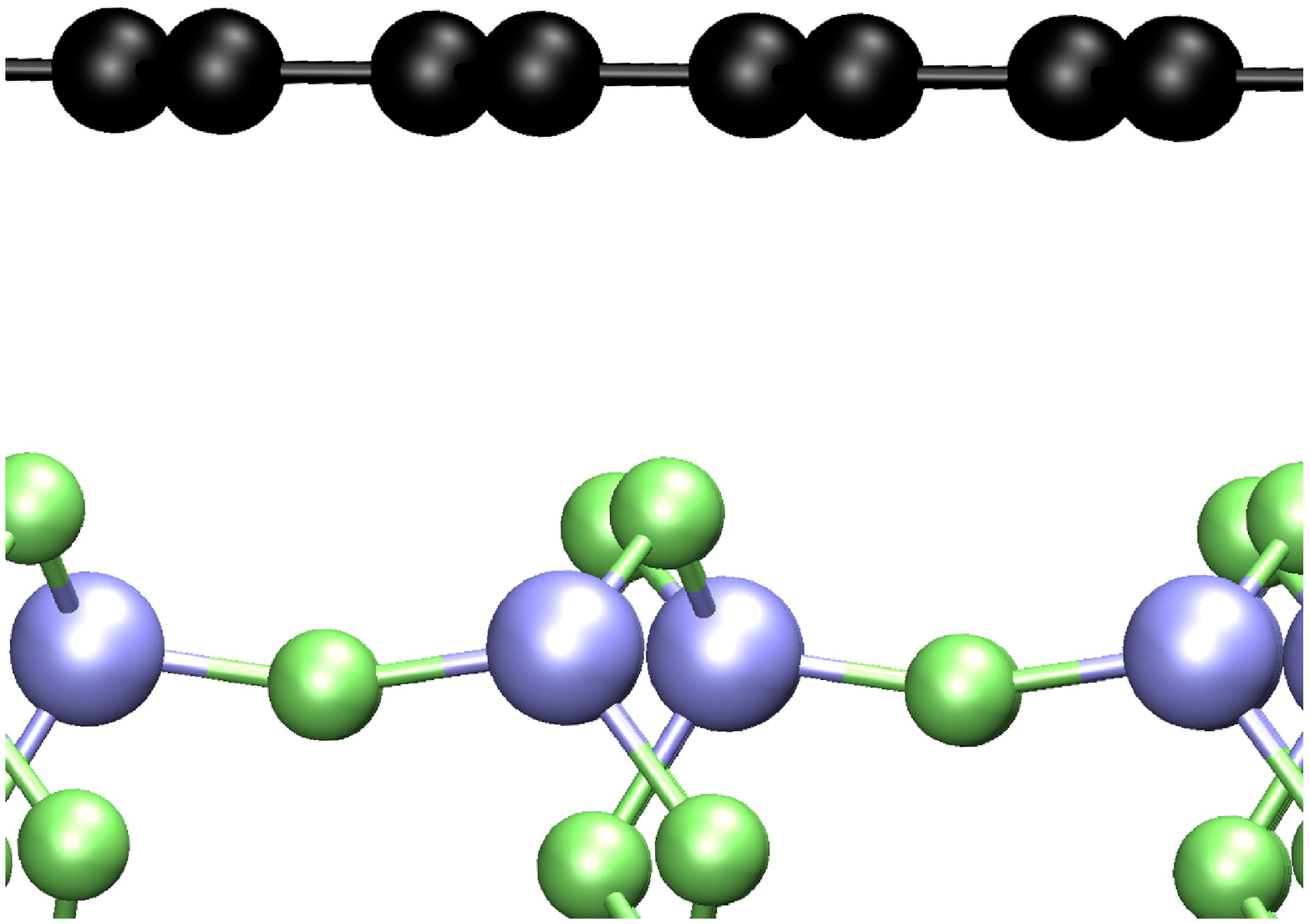}}\\ 
\end{tabular}

 \caption{\label{fig:graphene} The band structure along the $\Gamma$-K-K'-$\Gamma$ direction, and the geometry of graphene on SiO$_2$ with different surface terminations. OHT: hydroxyl terminated, OT: oxygen terminated, SiT: silicon terminated, ROT: reconstructed oxygen terminated. In the band structures, the dashed line shows the band structure of freestanding graphene, and in the geometries the computational supercell. Atom colors: Black -- carbon, blue -- silicon, green -- oxygen, yellow -- hydrogen.}
\end{figure*}

\section{Graphene on $\mathrm{\textbf{SiO}}_2$ surface}

For the relaxed geometries, we find only weak binding of graphene on all surfaces. Throughout this paper, with strong binding we refer to structures with substrate-carbon bonds, as opposed to weak binding characterized by longer substrate-carbon distances. The graphene layer is farther away from the substrate for the SiT surface (3.4~\AA{}), compared to the other surfaces ($\approx$3~\AA{}).\cite{OHTdist} The relative position of the graphene layer on the four different surfaces is presented in Fig.~\ref{fig:graphene}. For the SiT, ROT, and OHT surfaces, we get the hollow adsorption side, which fully agrees with previous work.\cite{Hossain, Kang, Cuong-Otani-Okada} For the OT surface, the uppermost oxygen atom of the substrate is slightly displaced from a bridge position between two carbons towards the center of the hexagon. Previous simulations have suggested top \cite{Hossain, Kang} or bridge \cite{Ao}  positions.

Otherwise, previous DFT studies have not found agreement on whether graphene strongly binds to the SiO$_2$ surface,\cite{Shemella-Nayak, Kang} or not.\cite{Cuong-Otani-Okada,Ao}  This discrepancy can also be seen in the values for the equilibrium distance of graphene from the substrate that range from 1.43~\AA{}\cite{Shemella-Nayak} to approximately 3.4~\AA{}.\cite{Kang}  During our relaxations, we find intermediate geometries with small forces that still continue to relax to clearly lower-energy geometries through slow sliding of the carbon layer on the surface. Because some of the intermediate states were bound to the surface, one possible cause for the discrepancy is the effect of numerical settings in the different calculations. At the very least, the relaxations need to be done with a high numerical accuracy, relaxing the forces until they are very small. Also the inclusion of the van der Waals correction can, in principle, affect the results but it should provide only additional attraction in comparison to the non-vdW case. 

The opening of the gap is connected to how tightly the graphene is bound to surface. In our calculations the gaps are zero, apart from the OT surface where we find a gap of approximately 10~meV. The band structure of graphene on all different surfaces shows the linear dispersion near the K-point, characteristic for freestanding graphene (see Fig.~\ref{fig:graphene}.).  Some substrate bands cross the graphene bands but only on the OT and OHT surfaces they are near the Fermi energy.  In the previous calculations on the OT surface, either the linear dispersion without a gap is preserved \cite{Cuong-Otani-Okada, Ao} or the graphene layer binds strongly, leading to a large gap.\cite{Kang, Shemella-Nayak} On the ROT and OHT surfaces, a small gap of approximately 20~meV has been reported.\cite{Cuong-Otani-Okada} In the experiments, the linear dispersion has been observed for SiO$_2$-deposited graphene.\cite{Zhang-Brar} \\  

\begin{figure*}
 \begin{tabular}{ccc}
\includegraphics[width = 0.325\textwidth]{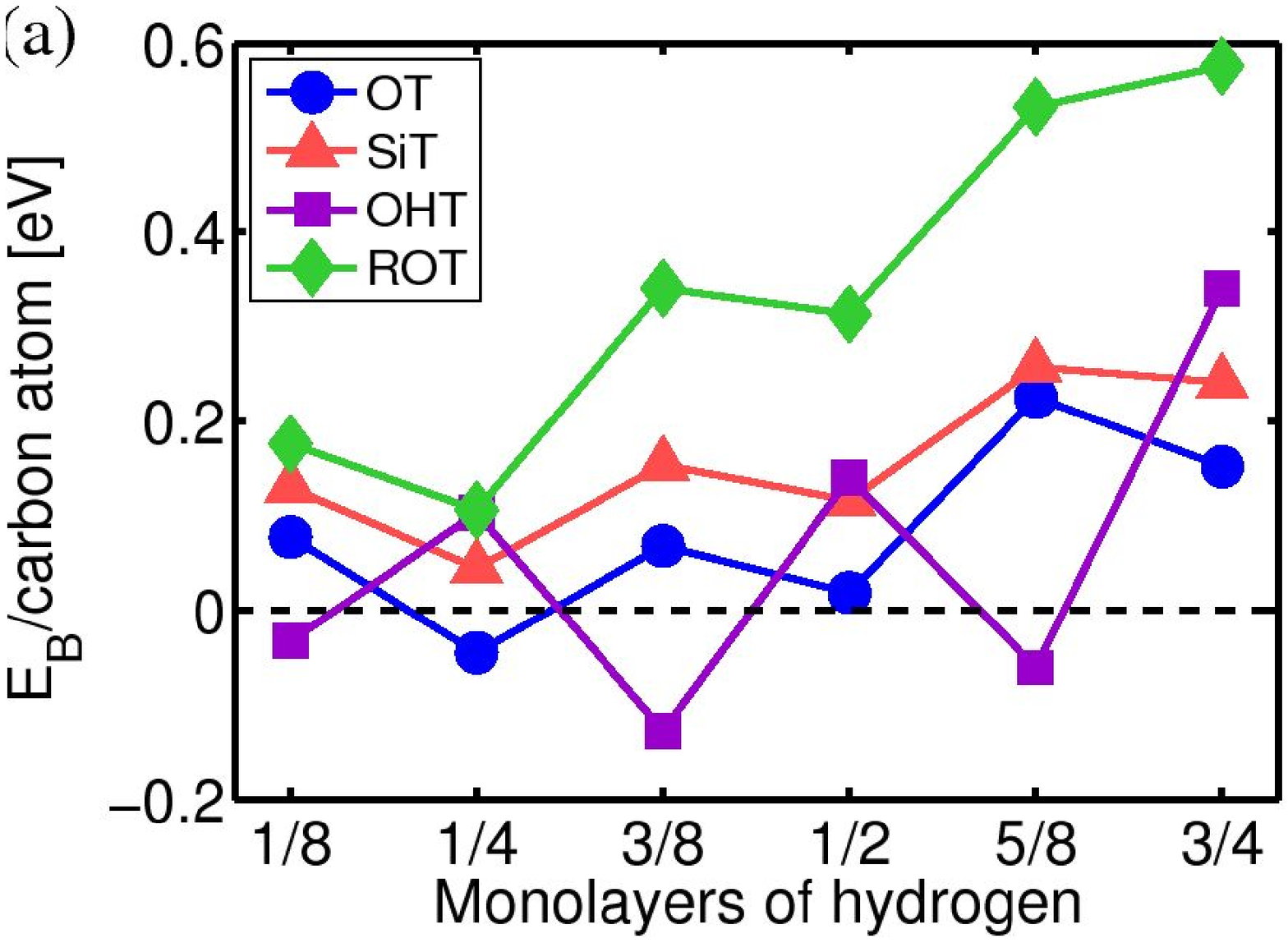} &
 \includegraphics[width= 0.325\textwidth]{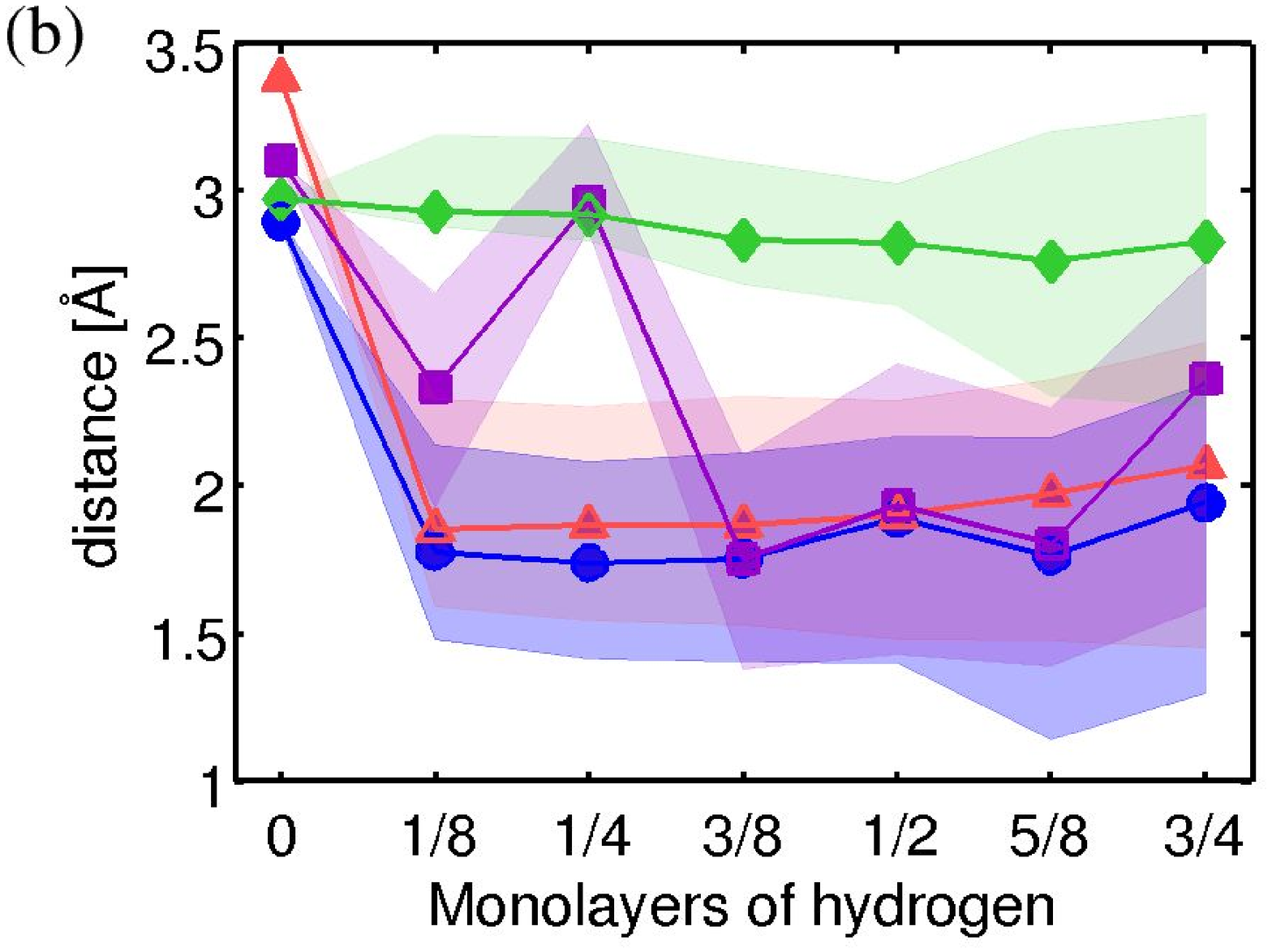}&
 \includegraphics[width = 0.325\textwidth]{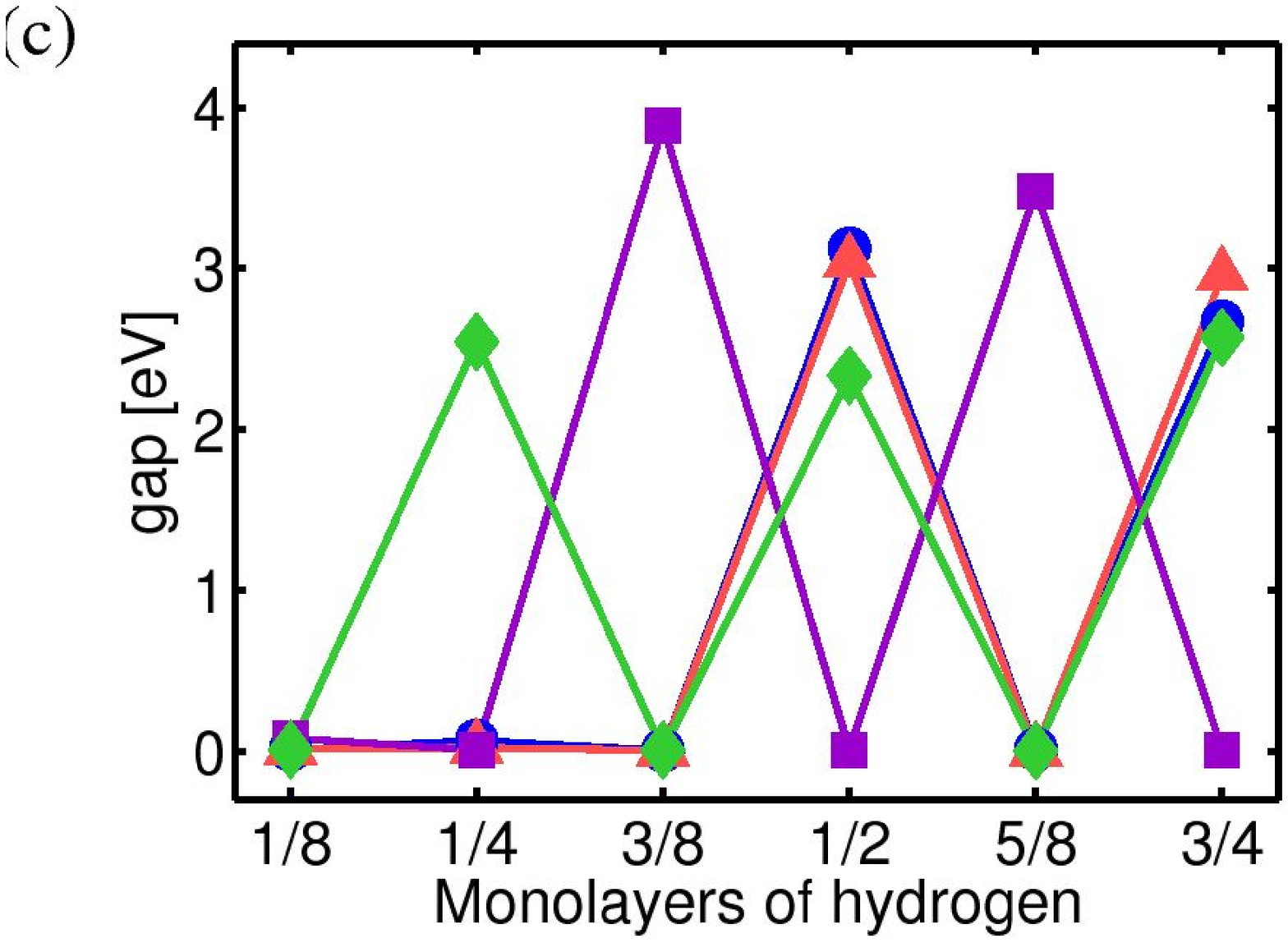}\\
 \end{tabular}
\caption{\label{fig:gap} (a) The binding energy (b)   Distance of the carbons from the substrate (c) Band gap as a function of the hydrogen content for the lowest-energy configurations for the four surface terminations. In (b), the shaded regions show the distance range and the markers the average distance. }
\end{figure*}

\section{Graphane on $\mathrm{\textbf{SiO}}_2$ surface}

When hydrogen atoms adsorb to the carbon sheet, the conducting graphene is transformed into a semiconducting graphane. In the freestanding fully hydrogenated graphane sheet, hydrogens attach to the carbons of different sublattices on different sides of the sheet.  The estimates for the semiconducting energy gap depend on the level of theory, ranging from approximately 3~eV of DFT to 6~eV in GW calculations.\cite{Sofo, Zhou-Wang, Bhattacharya, Lebegue, Leenaerts}  The carbon-carbon bond length in graphane, 1.5-1.6~\AA{},\cite{Sofo, Zhou-Wang, Bhattacharya, Lebegue, Leenaerts, Artyukhov} is longer than in graphene, 1.42~\AA{}. The longer bonds are counterbalanced by the non-planar structure of the carbon layer.  Using a pristine graphene sheet and molecular hydrogen as the reference state, we were not able to find a stable structure for freestanding one-sided hydrogenation of graphane. This is no surprise, as the freestanding one-sided graphane is a highly polar structure. In order to find out whether the surface can balance the polarity and make the structure stable, we study the one-sided hydrogenation of graphene on the four different SiO$_2$ surfaces.  

First, the graphane hydrogen configurations on the four different SiO$_2$ surfaces were determined. One to six hydrogen atoms corresponding to 1/8--3/4 ML coverage of the graphene carbon atoms were placed in the supercell on top of the carbon atoms. Then the carbon layer was deposited either on a top or hollow position with respect to the uppermost substrate atoms on a thinner slab corresponding to one unit cell of SiO$_2$ (5.3~\AA{}). The use of the thin slab allowed us to scan all different hydrogen initial configurations for both top and hollow positions, approximately 500 for each of the surfaces. The carbon layer and the hydrogens were relaxed. During the relaxation, a large number of local energy minima were found and a tight convergence criterion proved to be extremely important in order to locate the lowest-energy structures. Afterwards, the slab was increased to three SiO$_2$ unit cells (15.8~\AA{}) for the lowest-energy configurations on each surface and hydrogen coverage, and the three uppermost atom layers of the substrate were additionally relaxed.  For the calculation of the binding energy, the reference state was chosen to be the optimized structure for graphene on the substrate, along with a corresponding number of hydrogen molecules in a vacuum. 

Because the reconstruction on the ROT surface extends deep to the substrate, we use the unhydrogenated optimized graphene-OT interface on the other side of the slab for the extensive configuration scan to decrease the computational effort.  Due to the unsymmetry of the surfaces, a dipole correction was used. Again, a large number of initial hydrogen configurations was considered but during the relaxation of the carbon layer and the hydrogen atoms, the highest-energy ones were discarded and relaxation was continued only for the low-energy configurations. Later, the geometries were symmetrized and seven surface atom layers were  relaxed.
                                                                                               
Fig.~\ref{fig:gap}a shows the binding energies of the lowest-energy configurations as a function of the hydrogen coverage. We see that a graphane layer with hydrogen only on one side has indeed stable structures on the SiO$_2$ surfaces, in contrast to the freestanding one-sided graphane. The surface termination, however, plays a profound role on the stability as there are no stable structures on the SiT and ROT surfaces. In general, the structures with an even number of hydrogen atoms per supercell are more stable than their odd-numbered counterparts, and we see an oscillatory behavior of the binding energy. This applies also to the OHT surface if one takes into the account the hydrogen atom originating from the hydroxyl group of the substrate below the graphane layer. As there are multiple stable structures on the OHT surface, we note that even a small amount of hydrogen below the carbon layer stabilizes the structure and facilitates hydrogen adsorption above the carbon layer, also leading to higher hydrogen coverages. 

In general, unlike graphene, graphane is strongly bound to the substrate. This is illustrated in Fig.~\ref{fig:gap}b, where the average distance between the carbon and uppermost substrate atoms,\cite{OHTdist} along with its variation,  is shown
as a function of the hydrogen content. On the reactive OT surface, the graphane layer is strongly corrugated and one stable structure (1/4~ML) is found. On the contrary, the unreactive ROT surface leads to long carbon-substrate distances and high binding energies. The structures formed on the SiT surface are quite similar to the OT surface but not as strongly bound, as the average carbon-substrate distance is slightly longer, an exception being the 1/2~ML case where the distances are equal. On the OHT surface, the hydrogen atom below graphane facilitates the corrugation of the carbon layer and due to formation hydrogen-carbon bonds does not, in general, make the carbon-substrate distance longer.\cite{OHTdist}

The band gap of graphane (Fig.~\ref{fig:gap}c) does not increase monotonously when hydrogen coverage is increased but it shows a similar oscillatory behaviour as the energetic stability (Fig.~\ref{fig:gap}a). The structures with large energy gaps are energetically more stable and have an even number of hydrogen atoms in the supercell (counting again 
one hydrogen atom below the carbon layer on the OHT surface). Some of the hydrogen coverages show no energy gap at all.  In most cases, this is due to relatively flat, almost nondispersive carbon bands at the Fermi level. It should be noted that DFT is known to underestimate band gaps and in the case of freestanding graphane, the use of the GW method leads to 1.5--2 times larger gaps.\cite{Lebegue, Leenaerts}

\begin{figure*}

\begin{tabular}{ccc}
{\includegraphics[width=0.325\textwidth]{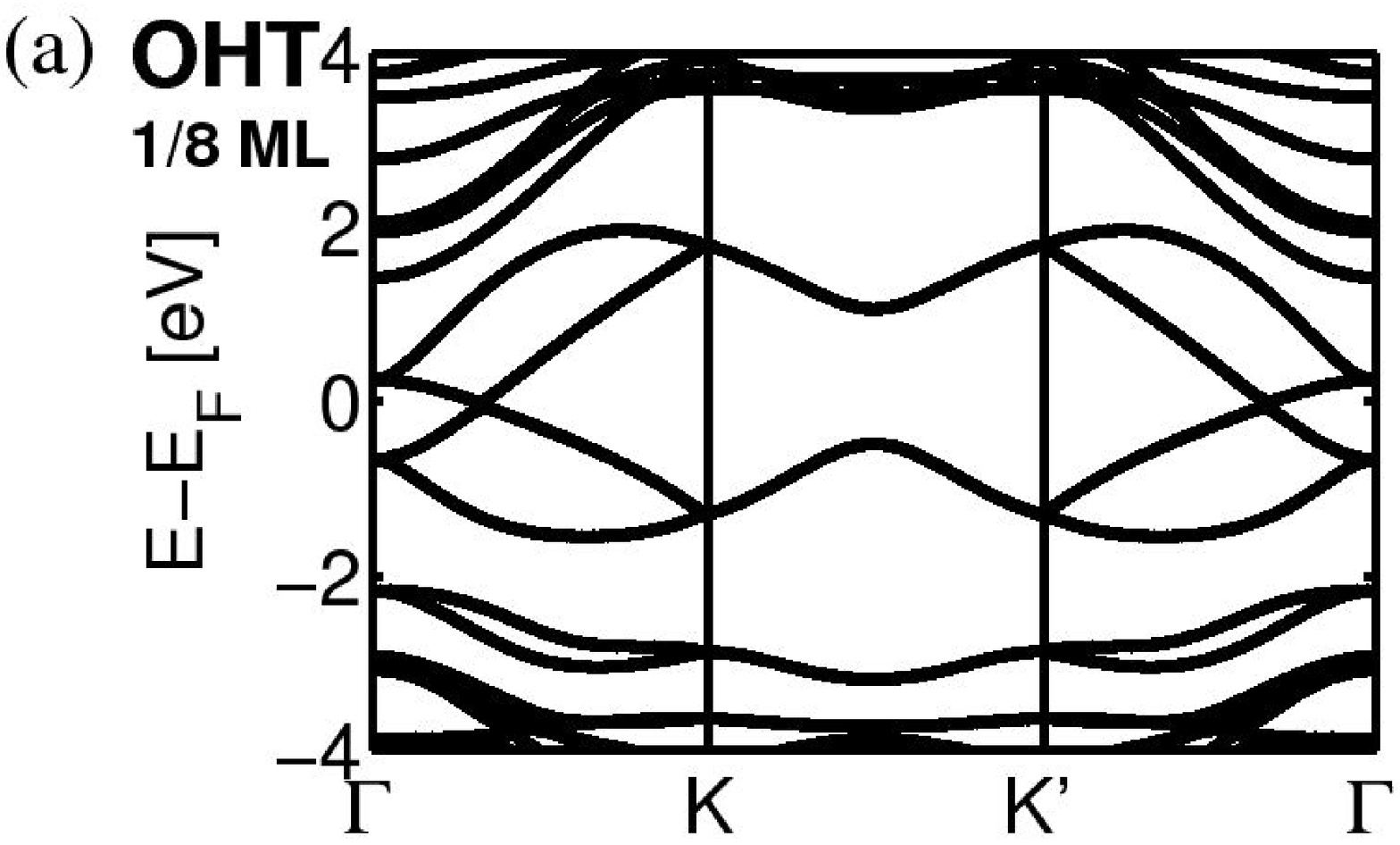}} & 
 {\includegraphics[width=0.325\textwidth]{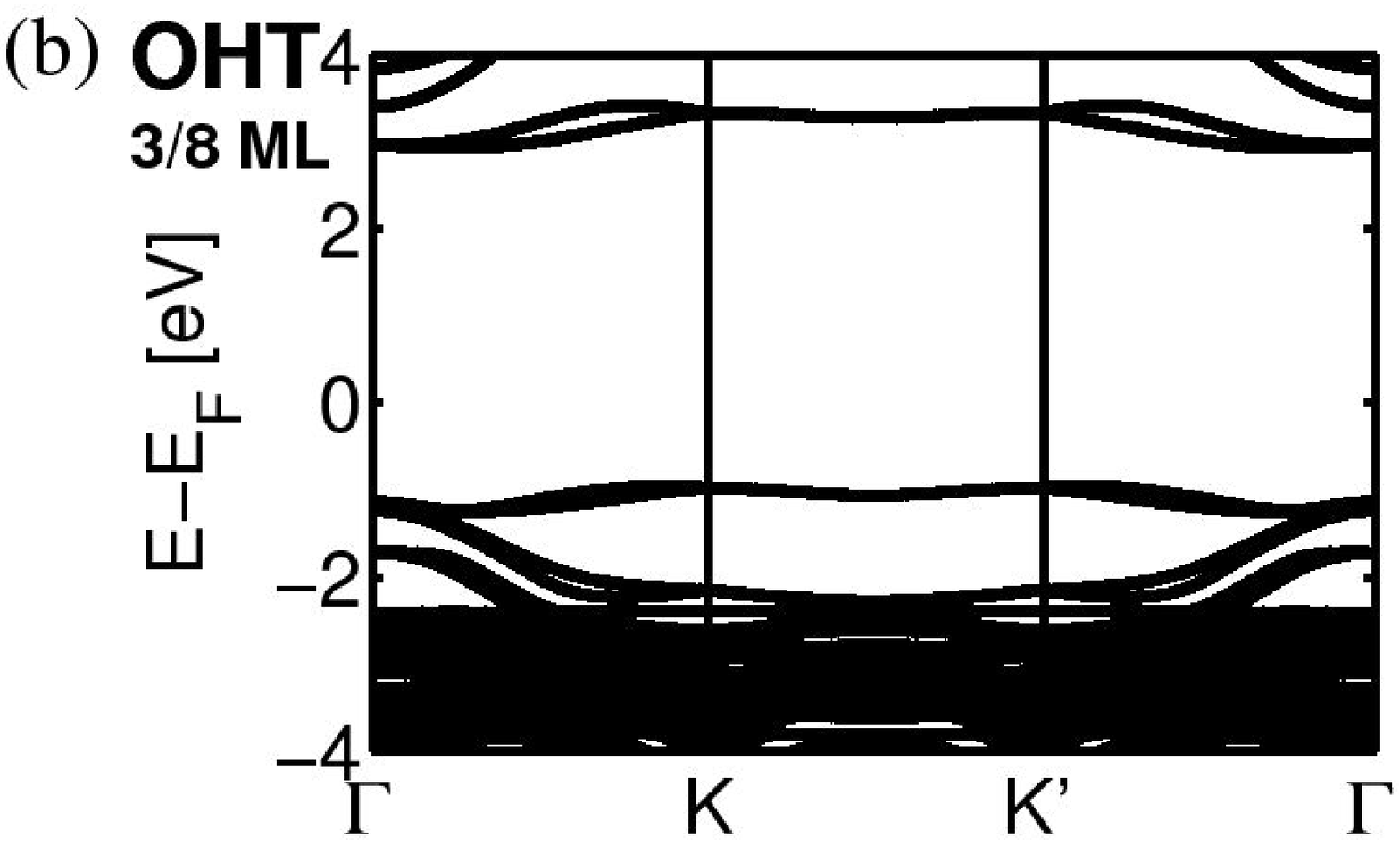}} & 
{\includegraphics[width=0.325\textwidth]{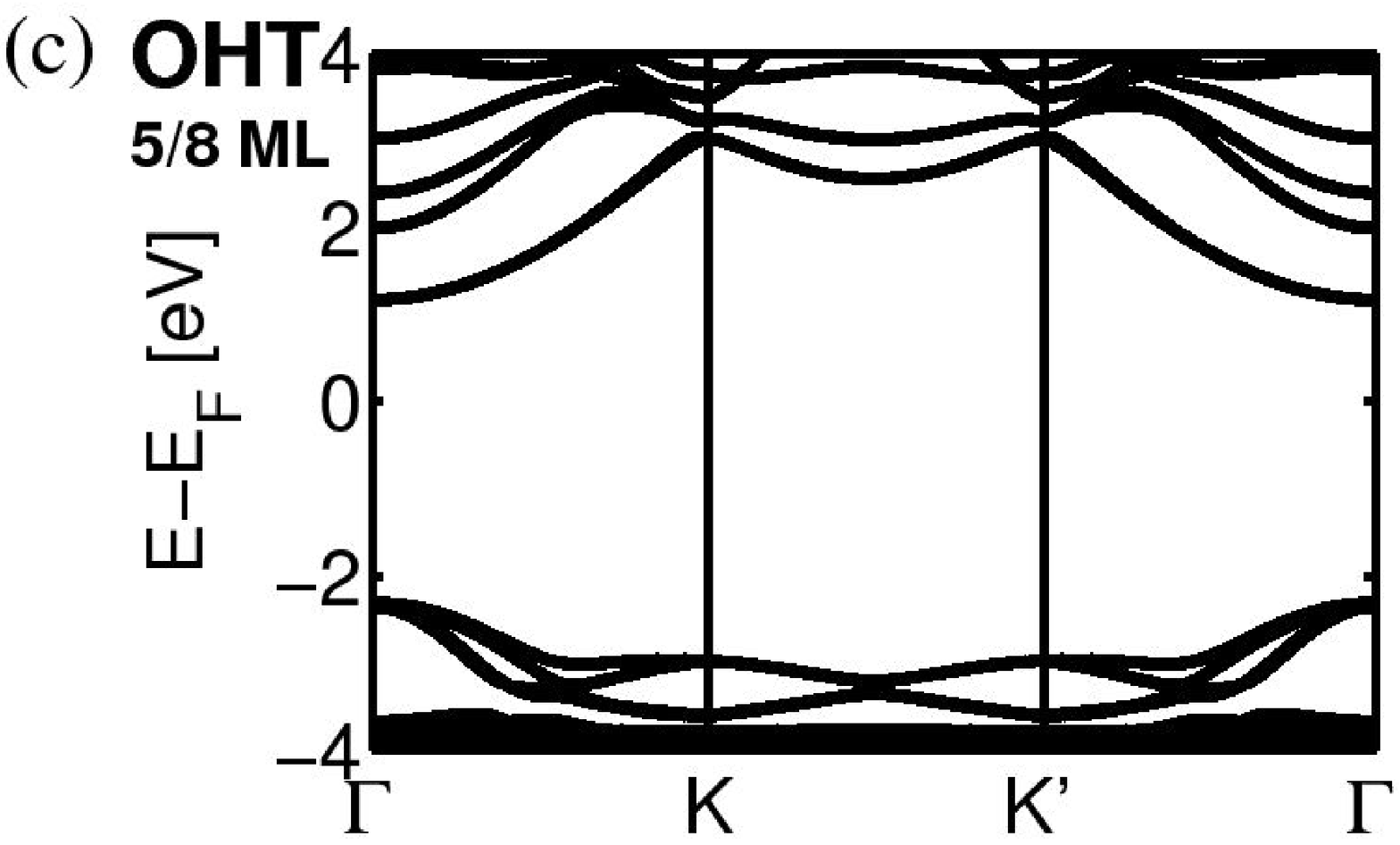}} \\ 
\hspace{0.07\columnwidth} {\includegraphics[width=0.22\textwidth]{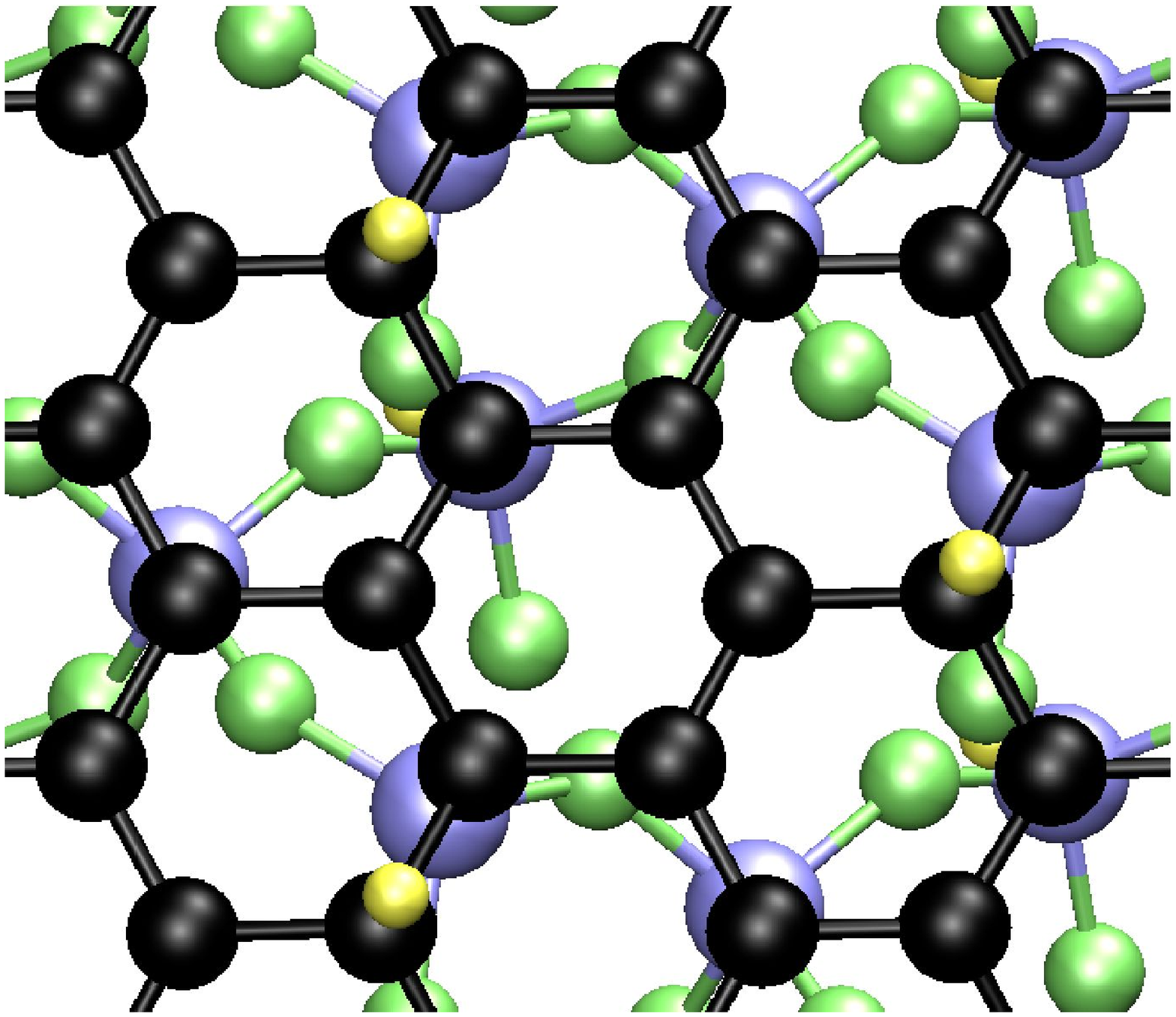}} & 
\hspace{0.07\columnwidth}{\includegraphics[width=0.22\textwidth]{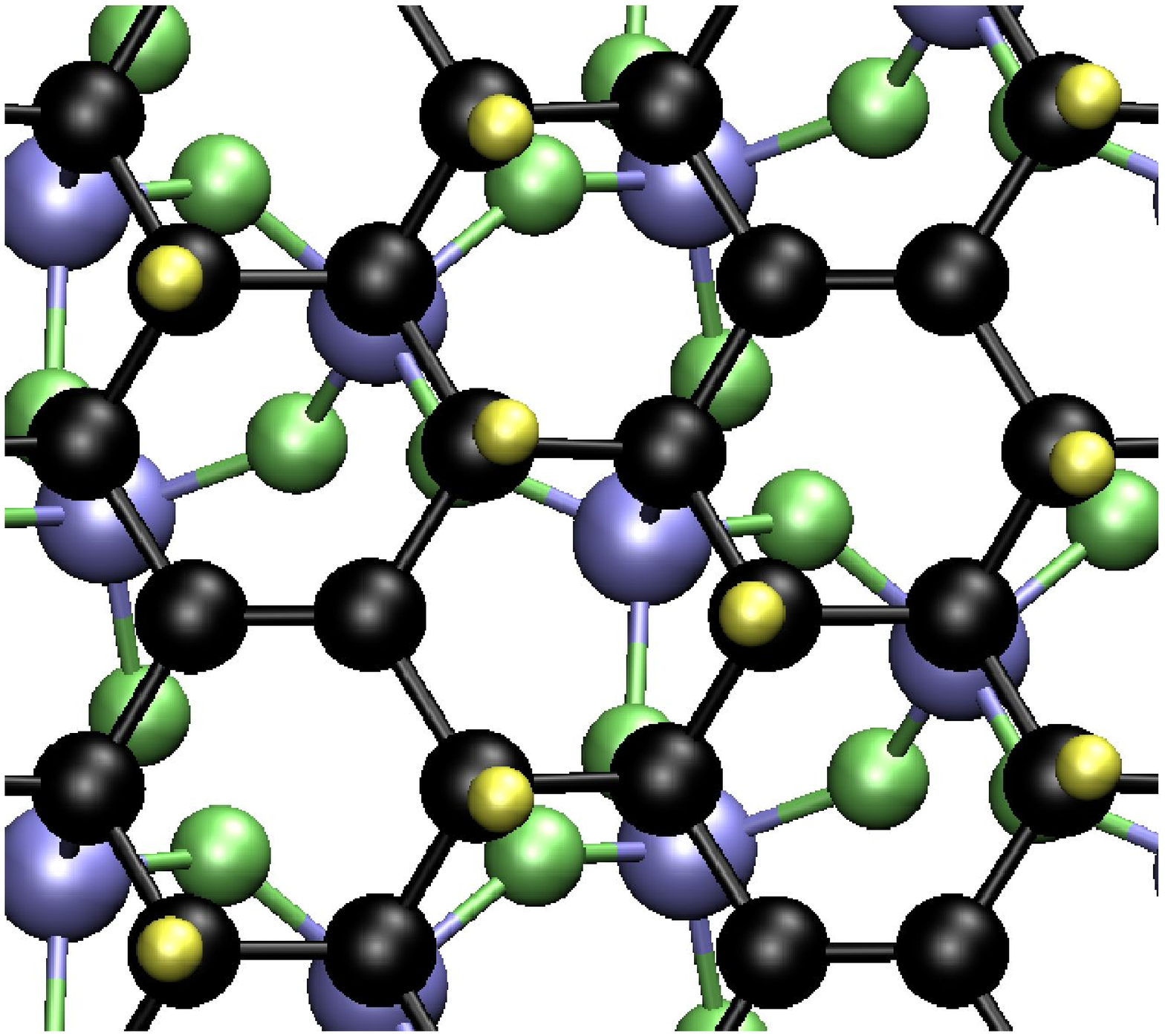}}& 
\hspace{0.07\columnwidth}{\includegraphics[width=0.22\textwidth]{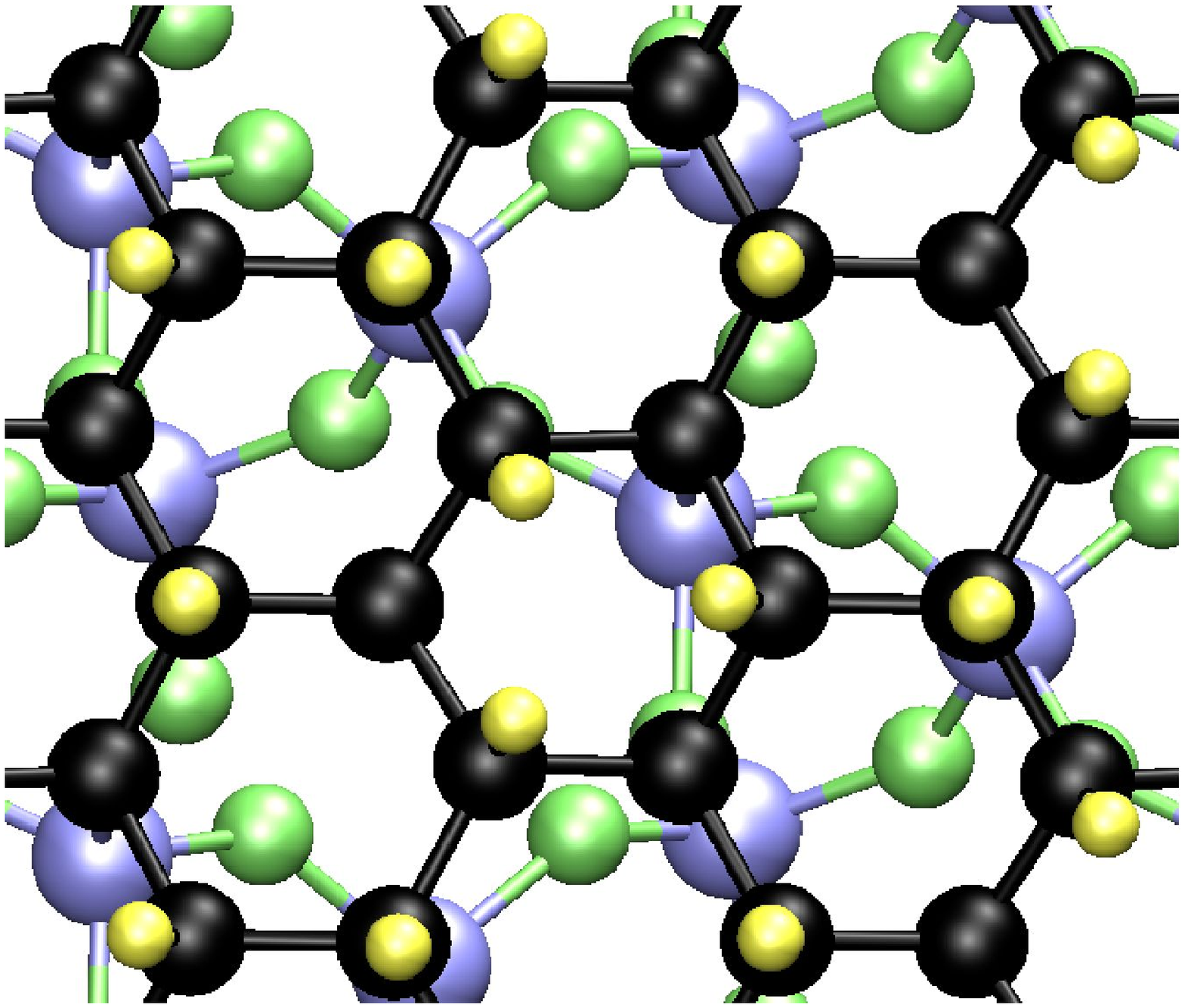}} \\ 
\hspace{0.07\columnwidth}{\includegraphics[width=0.22\textwidth]{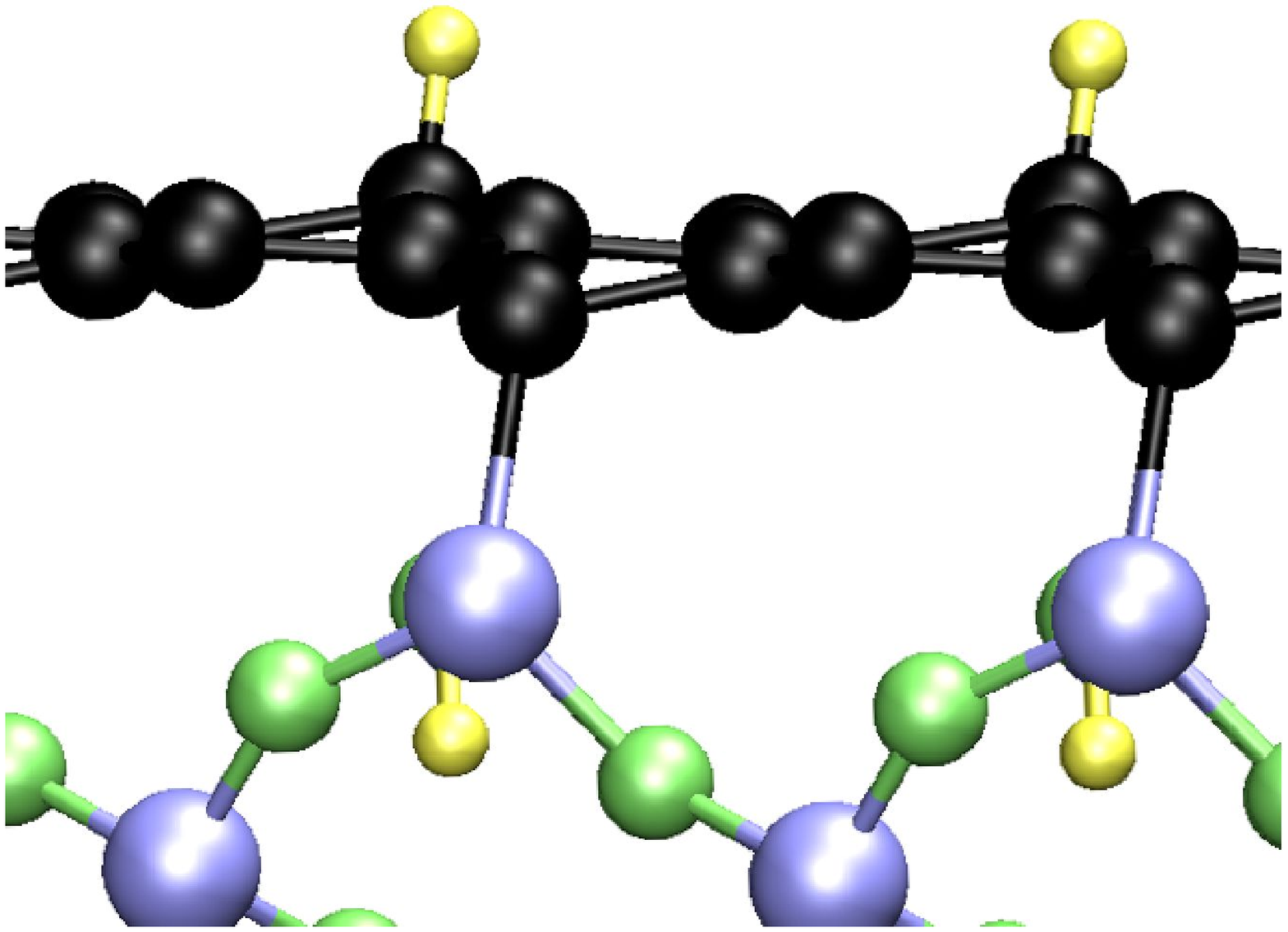}} & 
\hspace{0.07\columnwidth}{\includegraphics[width=0.22\textwidth]{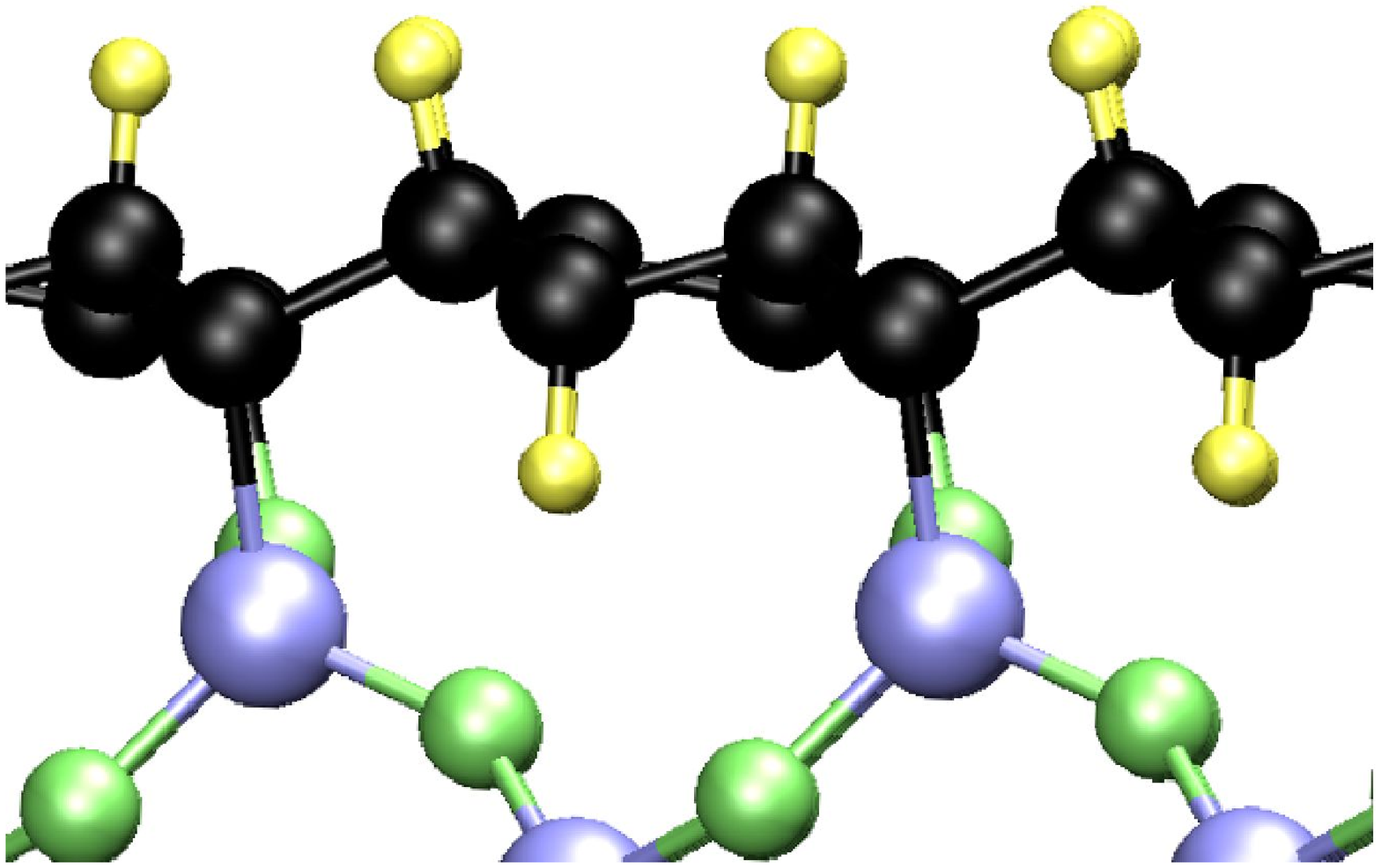}}&  
\hspace{0.07\columnwidth}{\includegraphics[width=0.22\textwidth]{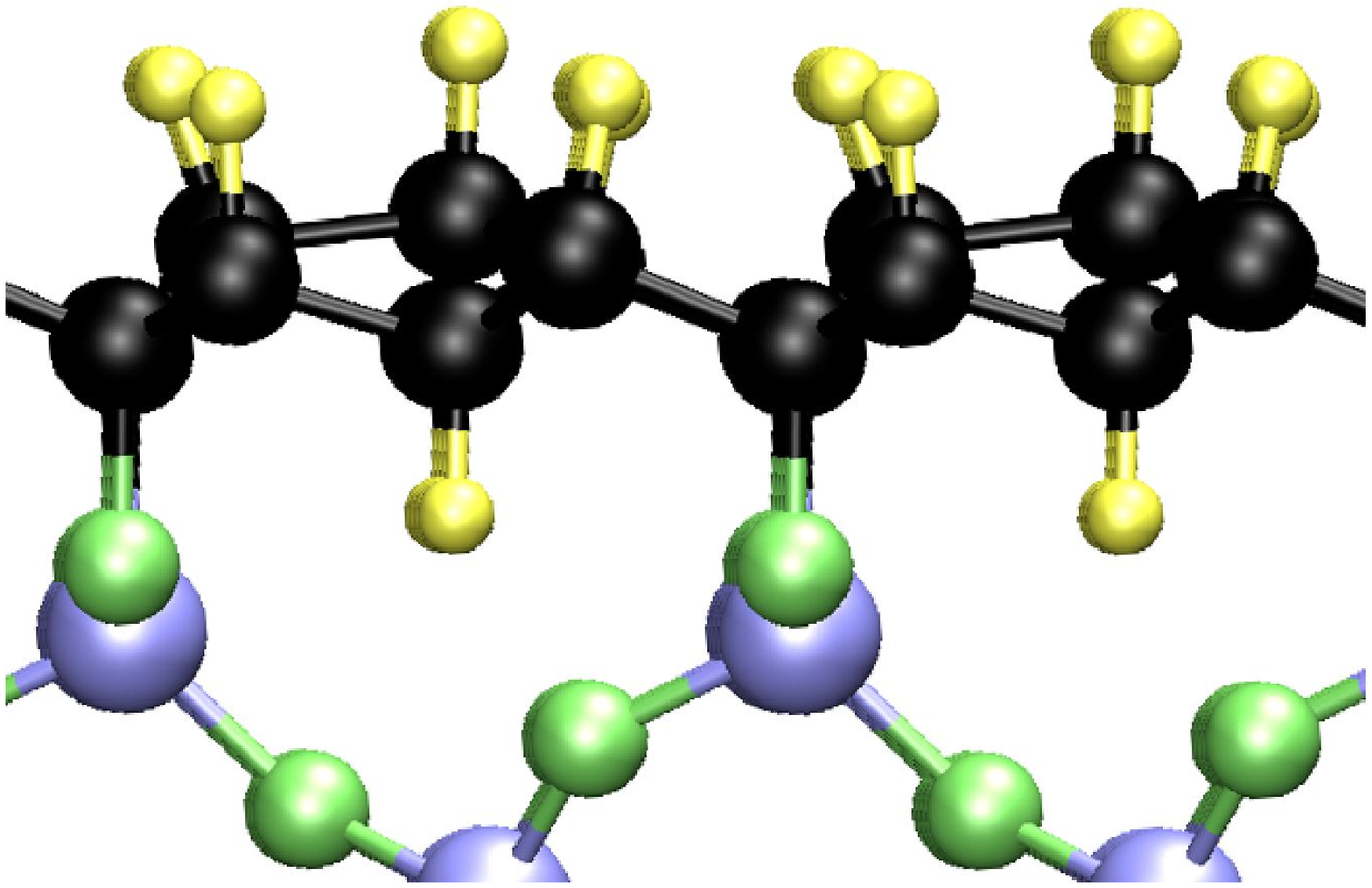}}\\ 

{\includegraphics[width=0.32\textwidth]{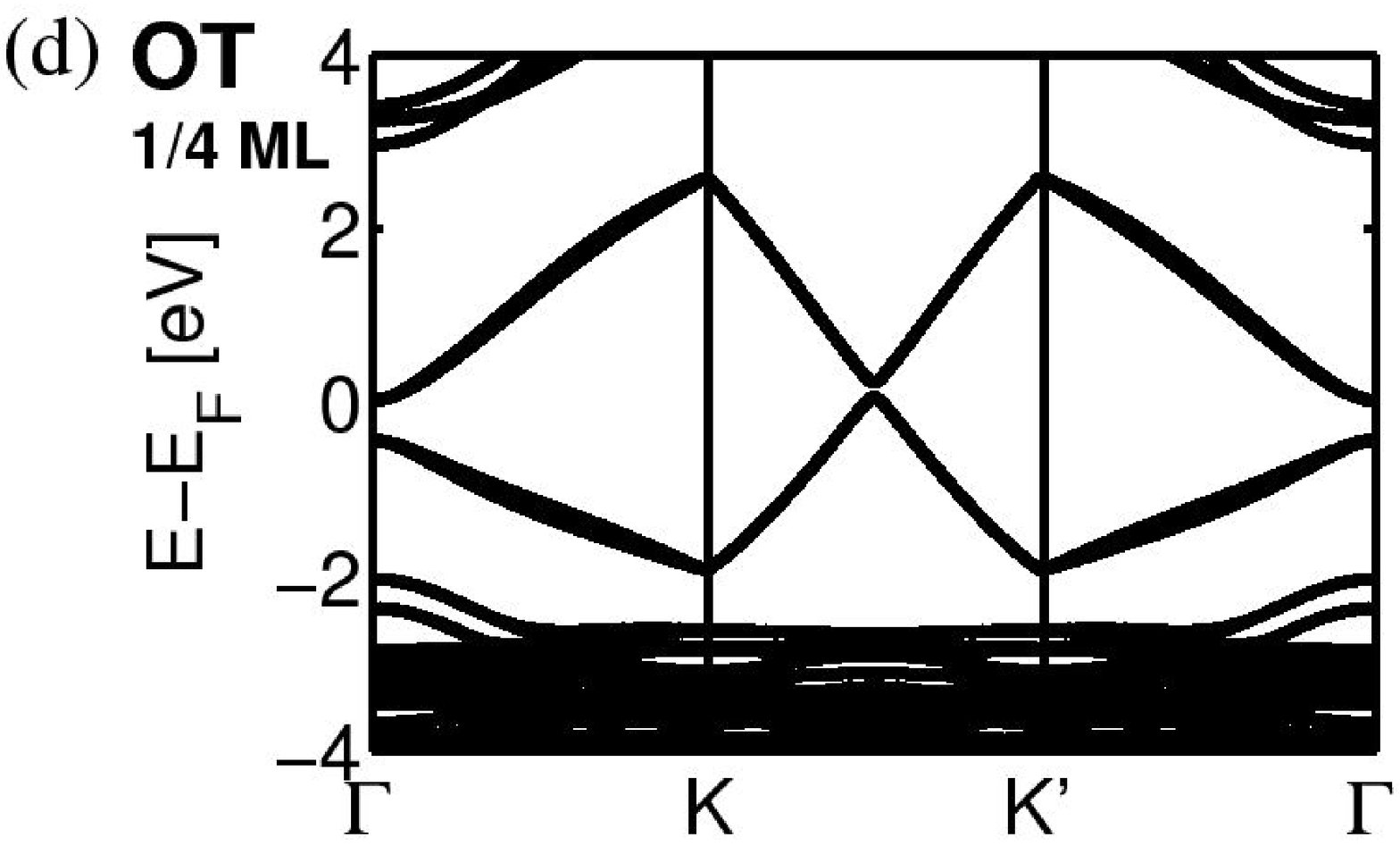}} &  
{\includegraphics[width=0.32\textwidth]{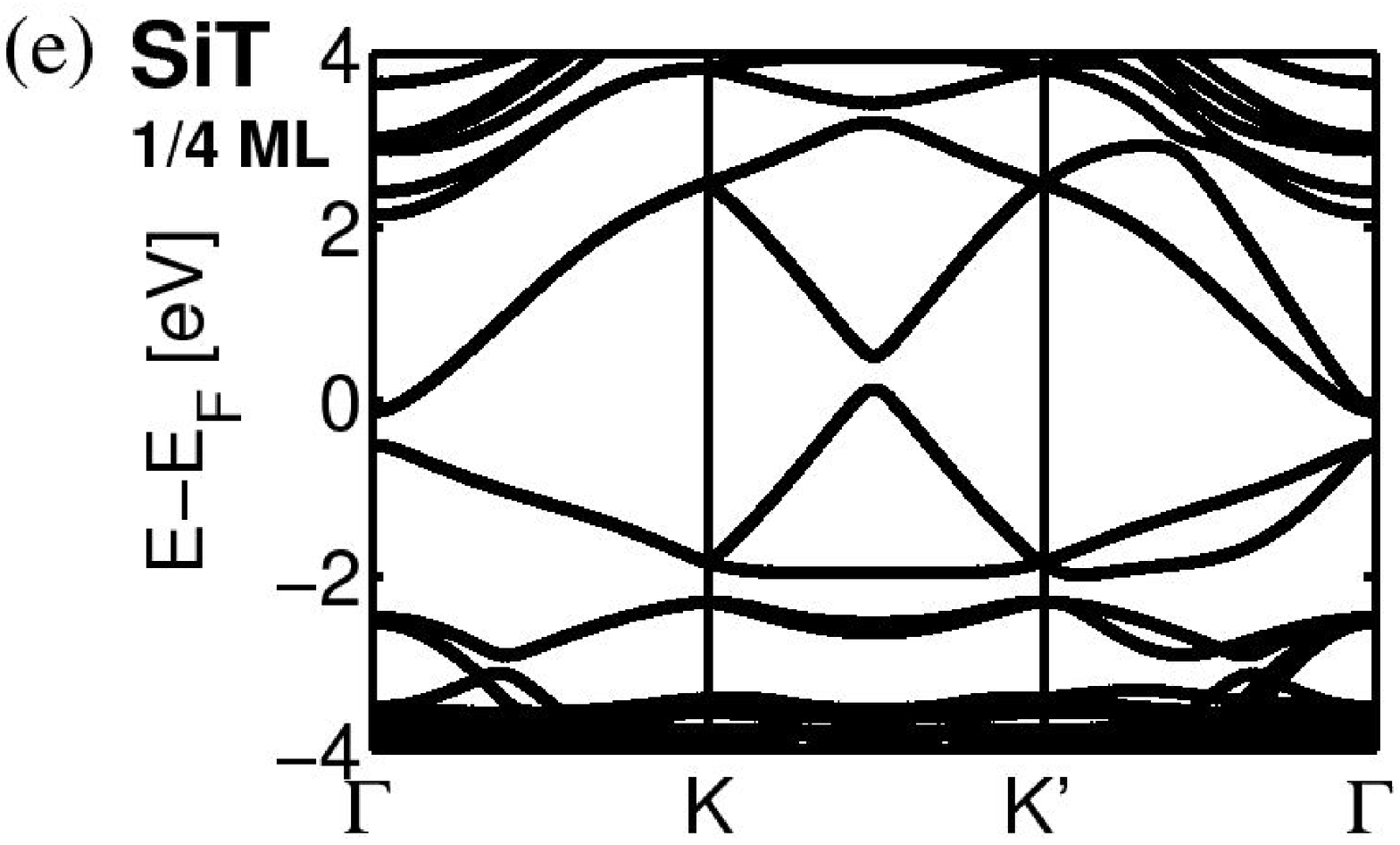}} & 
{\includegraphics[width=0.32\textwidth]{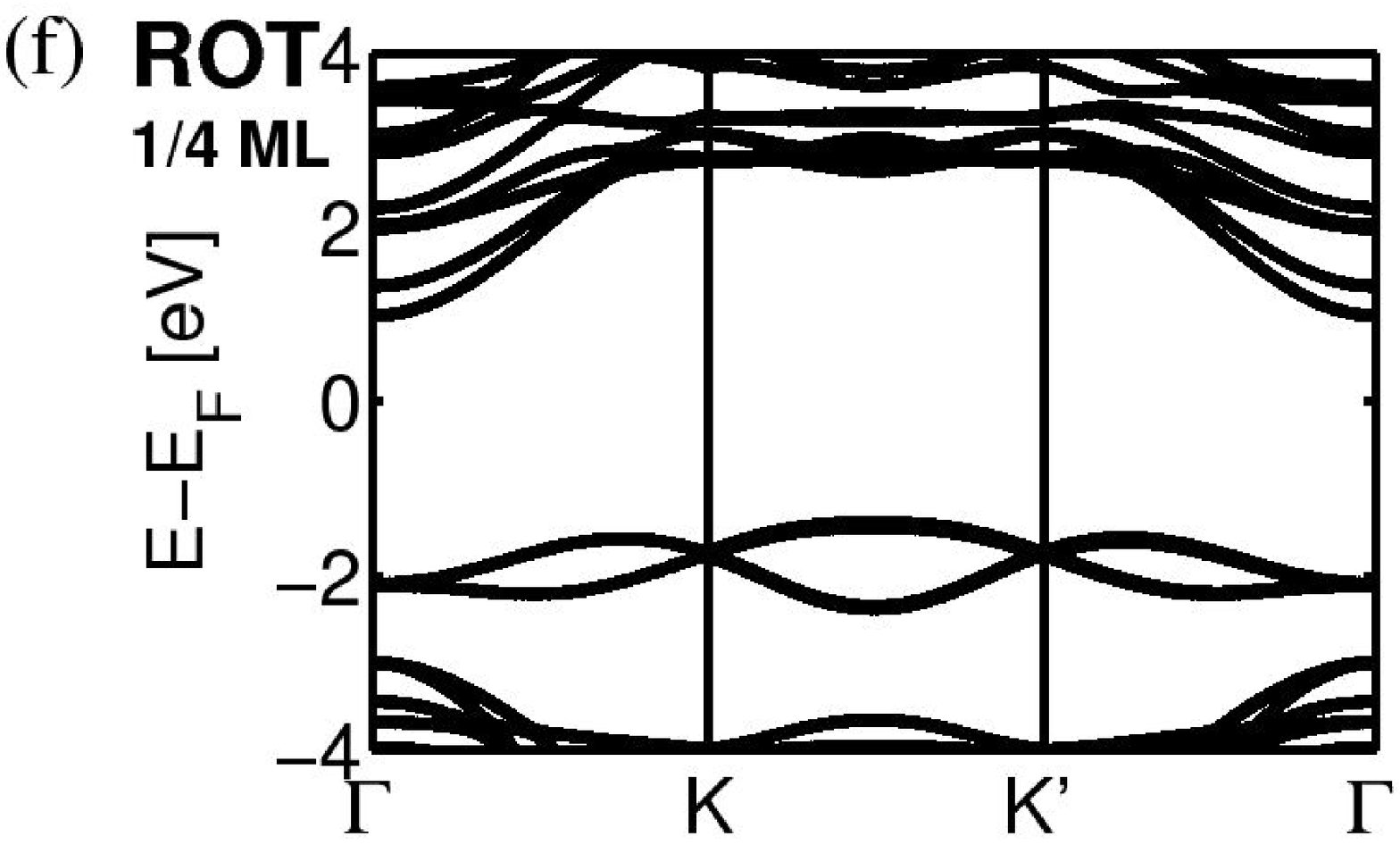}} \\ 
\hspace{0.07\textwidth}{\includegraphics[width=0.22\textwidth]{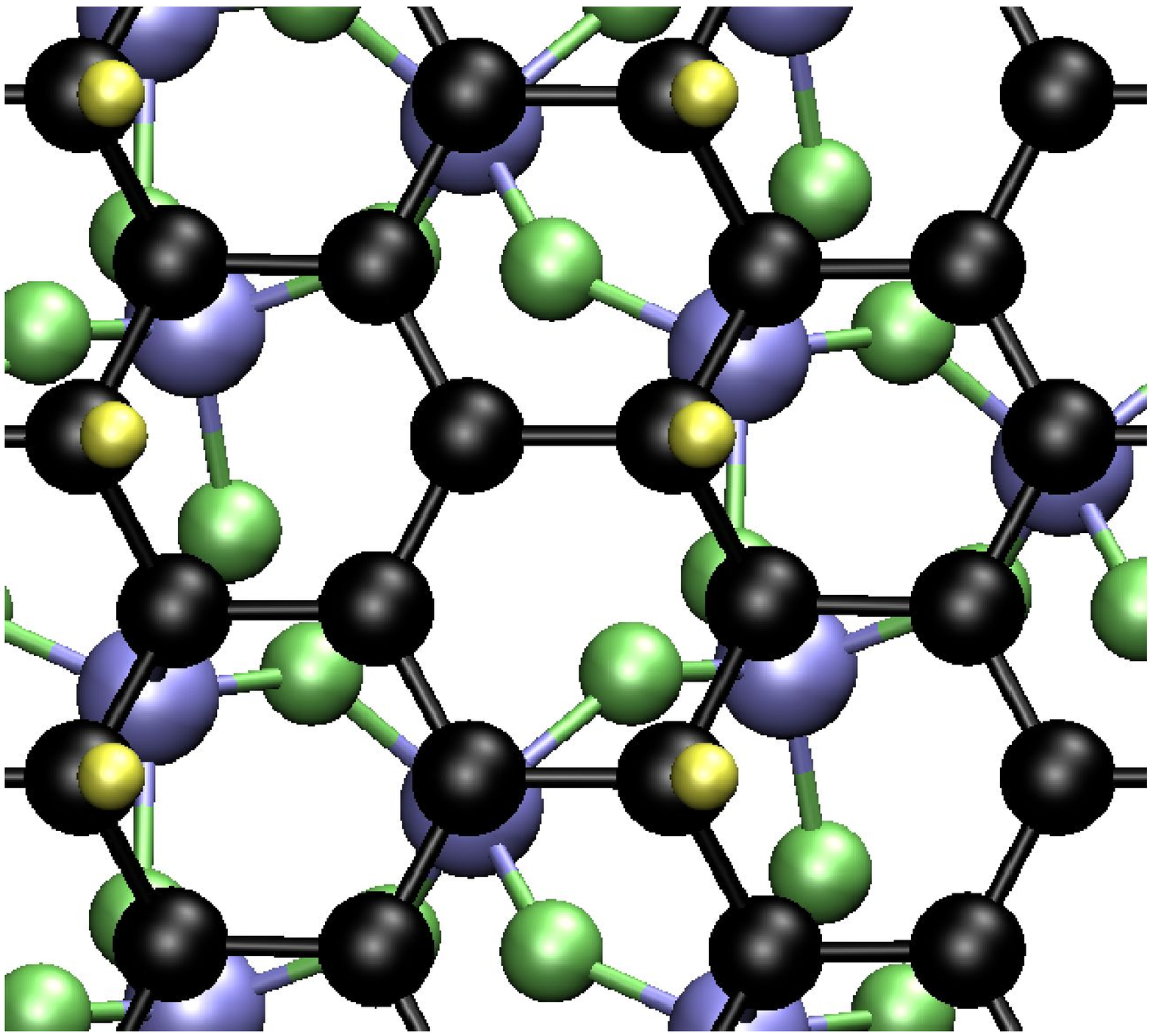}}& 
\hspace{0.07\textwidth}{\includegraphics[width=0.22\textwidth]{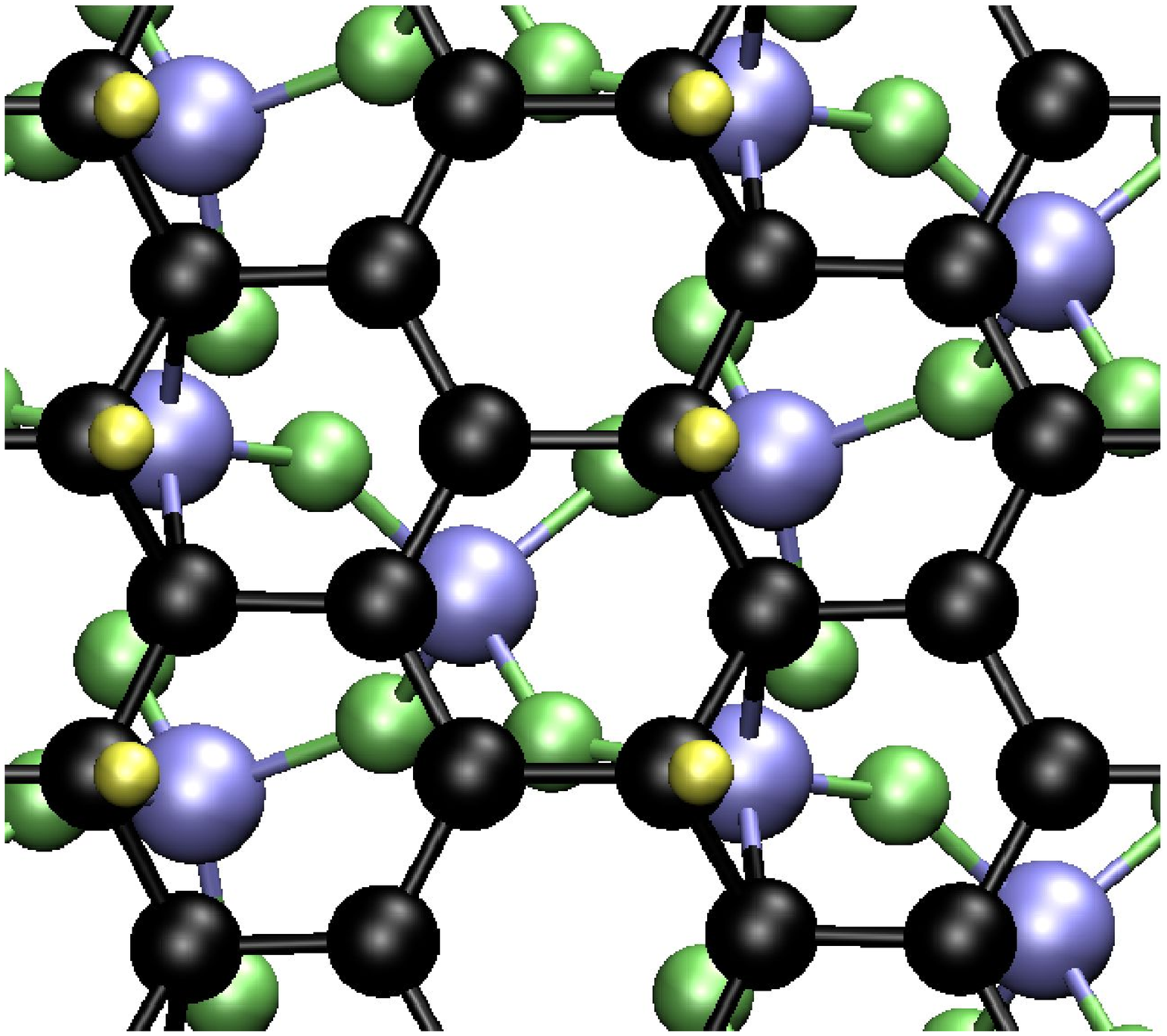}}& 
\hspace{0.07\textwidth}{\includegraphics[width = 0.22\textwidth]{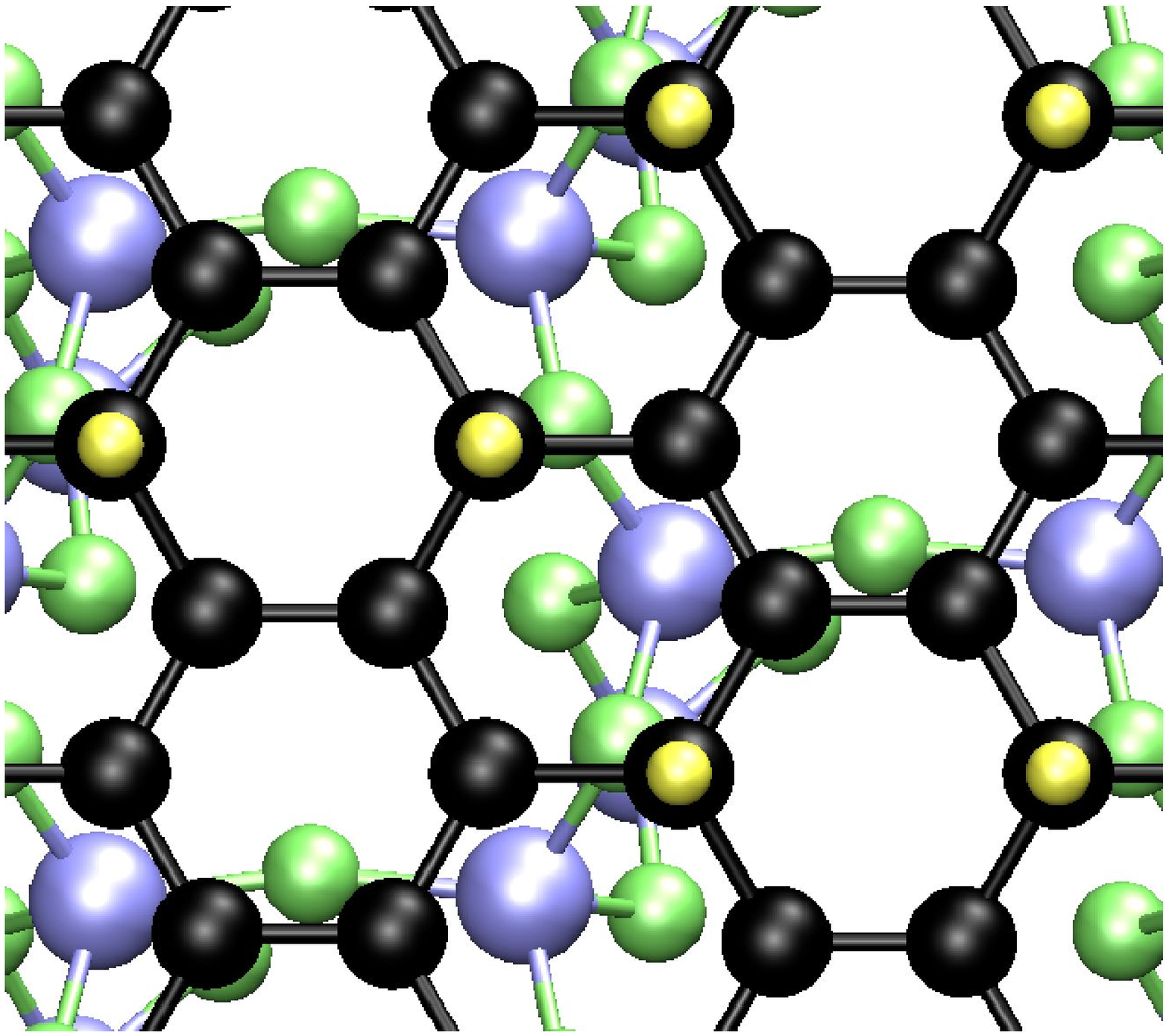}} \\ 
\hspace{0.07\textwidth}{\includegraphics[width=0.22\textwidth]{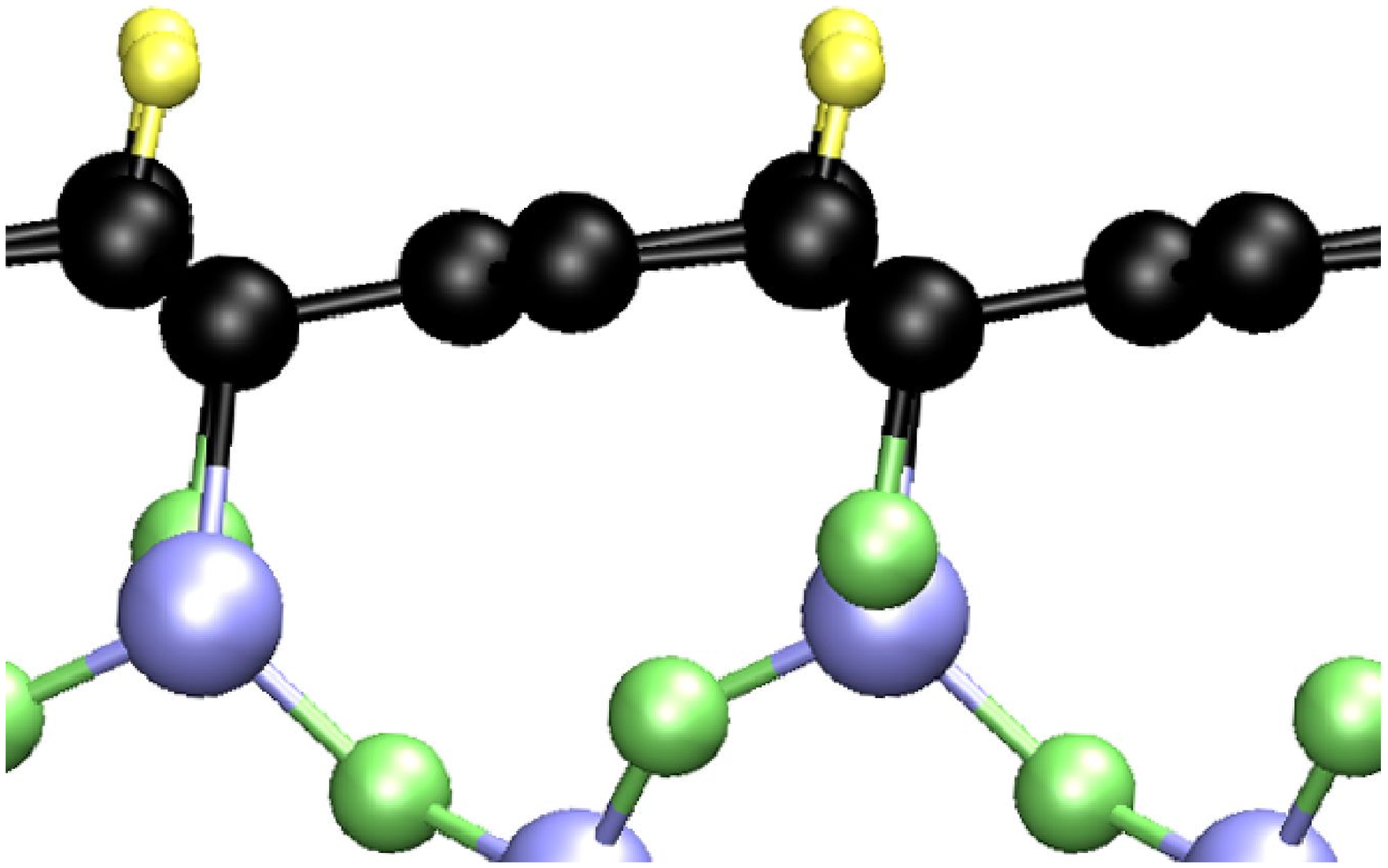}}&  
\hspace{0.07\textwidth}{\includegraphics[width=0.22\textwidth]{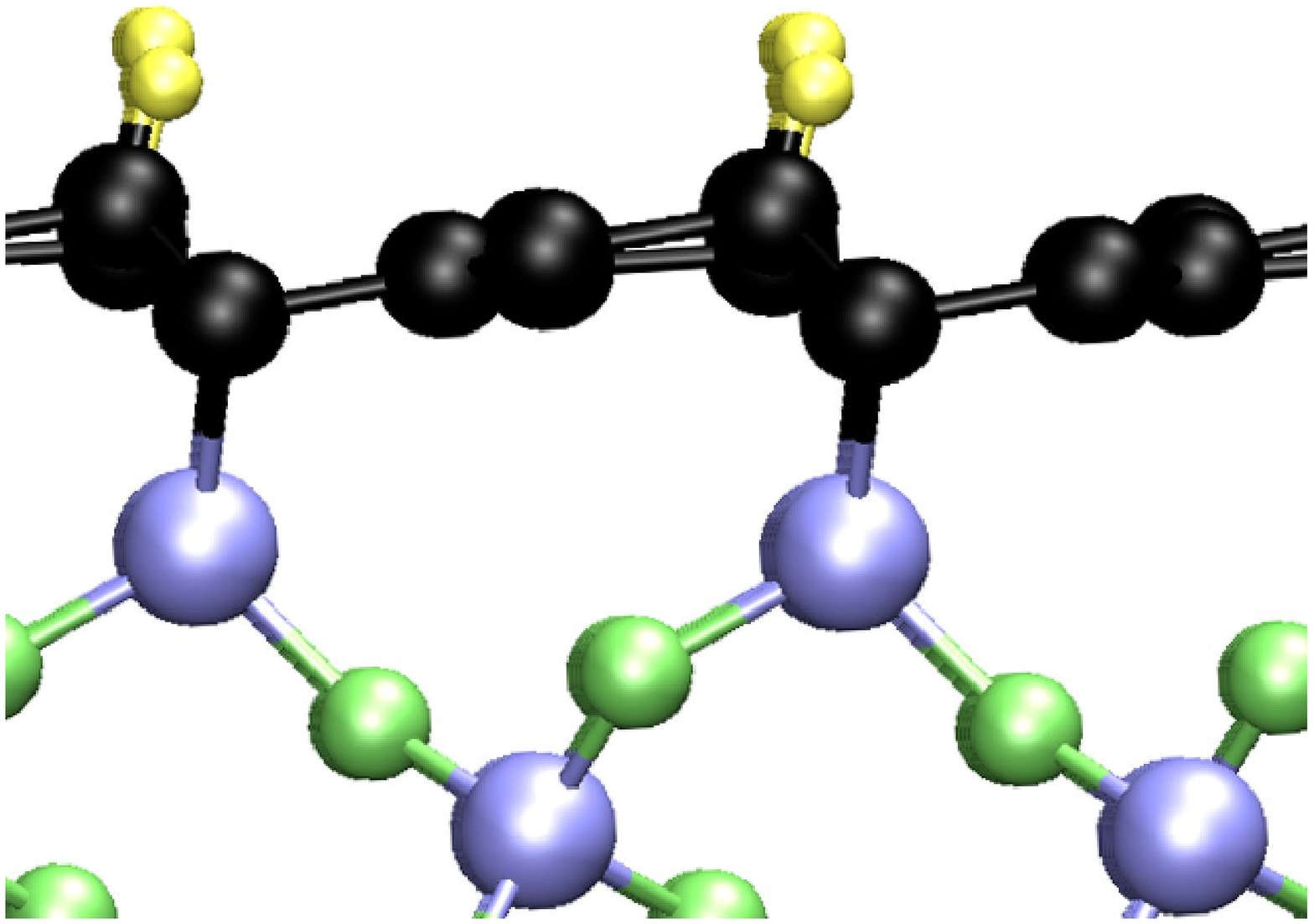}}& 
\hspace{0.07\textwidth}{\includegraphics[width = 0.22\textwidth]{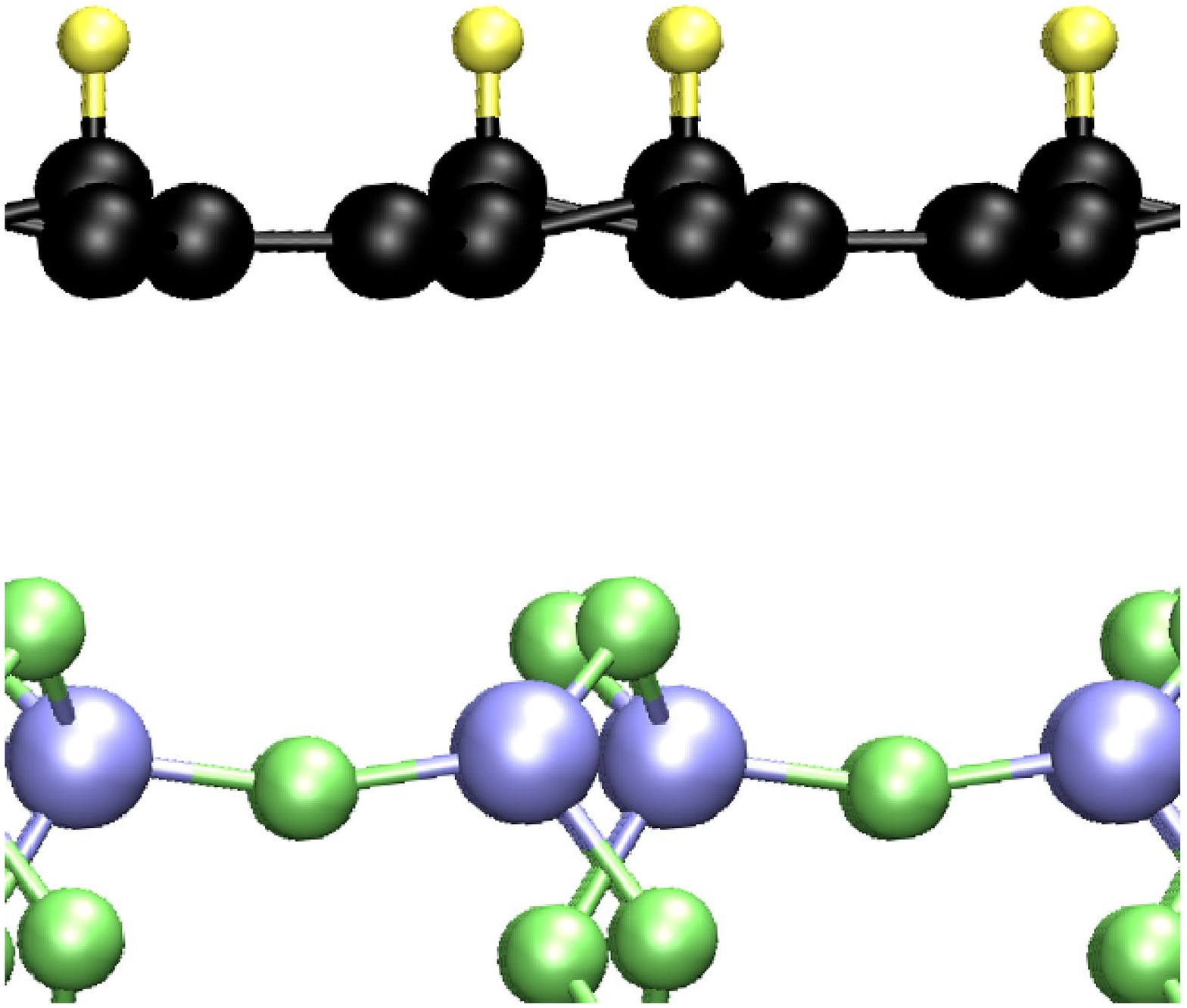}}\\ 

\end{tabular}

\caption{\label{fig:geoband} Band structure and hydrogen configuration for the lowest-energy graphane structures. OHT: (a) 1/8~ML coverage (b) 3/8~ML coverage (c) 5/8~ML coverage above the carbon layer.  1/4~ML coverage: (d) OT (e) SiT (f) ROT surface. Note that in the hydrogenated structures, the K and K' symmetry points are  no longer equivalent. In (d) and (e), unhydrogenated zigzag carbon lines with a Peierls instability are formed (see text for details). Atom colors like in Fig.~\ref{fig:graphene}. }
\end{figure*}

The three most stable structures on the OHT surface are, in the order of energetic stability, the 3/8, 5/8 and 1/8~ML hydrogen coverages above the carbon layer. The band structures and hydrogen configurations of these structures are shown in Fig.~\ref{fig:geoband}a-c. For the 1/8~ML coverage, the substrate hydrogen remains attached to the surface oxygen whereas for 3/8 and 5/8~ML coverages, the hydrogen atom binds from below to the carbon layer. The structure of these two states resembles that of freestanding graphane in the chair configuration as the hydrogen atoms above the carbon plane bind only (mostly) to one sublattice for the 3/8 (5/8)~ML structure, and most of the neighbours of these atoms bind either to the substrate or to a sub-carbon hydrogen.

Also the OT surface forms a stable structure at 1/4~ML hydrogen coverage.  In addition, the binding energy of graphane on the OT surface with 1/2~ML coverage is positive but close to zero, and thus the structure is possibly a metastable state. Hydrogenated structures with 1/4~ML coverage are lowest in energy also on SiT and ROT surfaces but their binding energies are significantly higher. 

On OT and SiT surfaces, the hydrogen atoms in the 1/4~ML configurations form rows along every  other graphene zigzag line, the carbon atoms of the hydrogenated lines alternatingly binding to a hydrogen atom above the carbon plane and to the substrate (see Fig.~\ref{fig:geoband}de).  Interestingly, the corresponding band structures show a single Dirac cone between the K and K'-points of the first Brillouin zone, instead of the two cones seen for graphene at K and K'. A gap of magnitude 0.2~eV and 0.45~eV on the OT and SiT surfaces, respectively, appears between the tips of the two cones , but the Fermi level is not within this gap, reducing the indirect gap to few meV. The lowest-energy hydrogenated structure on the ROT surface, on the contrary to the OT and SiT counterparts, is a semiconductor with a 2.5~eV band gap and does not show any significant directional variation in the band structure. The hydrogen atoms adsorb only to one sublattice but they do not form lines.

We gain more insight into the merging of the two Dirac cones of 1/4~ML OT and SiT structures by plotting the electronic bands both along the direction of the hydrogenated chains, and perpendicular to it. In the transverse direction, the bands are almost flat and nondispersive whereas along the direction of the zigzag chains, the Dirac cone is present. Effectively, the systems are thus one-dimensional conductors, although some coupling over the hydrogen lines through the substrate remains.  Very recently,\cite{Kim} a Peierls instability was found in similar unhydrogenated zigzag chains in freestanding graphane. The instability is seen also in our structure, with the bond length alternation along the zigzag chain being of somewhat smaller amplitude than in the freestanding structure ($\Delta d$ = 0.008-0.012~\AA{} compared to 0.015~\AA{}\cite{Kim}). This is expected, as the bond lengths of graphane on the substrate are generally compressed in comparison to the freestanding structure. 

The two graphene Dirac cones have been  suggested to be useful for valleytronics,\cite{Rycerz} an analog of spintronics using the valley isospin arising from the inequivalence of the two Dirac cones. Topological line defects have been suggested to give rise to valley filtering.\cite{Gunlycke} The valley degree of freedom is also thought to be robust against disorder.\cite{Rycerz} As there is only a single valley in the hydrogenated structure, our results suggest that the valley degree of freedom could be destroyed by ordered adsorption. \\

\section{Thermodynamical stability}

\begin{figure}
\includegraphics[width=0.47\textwidth]{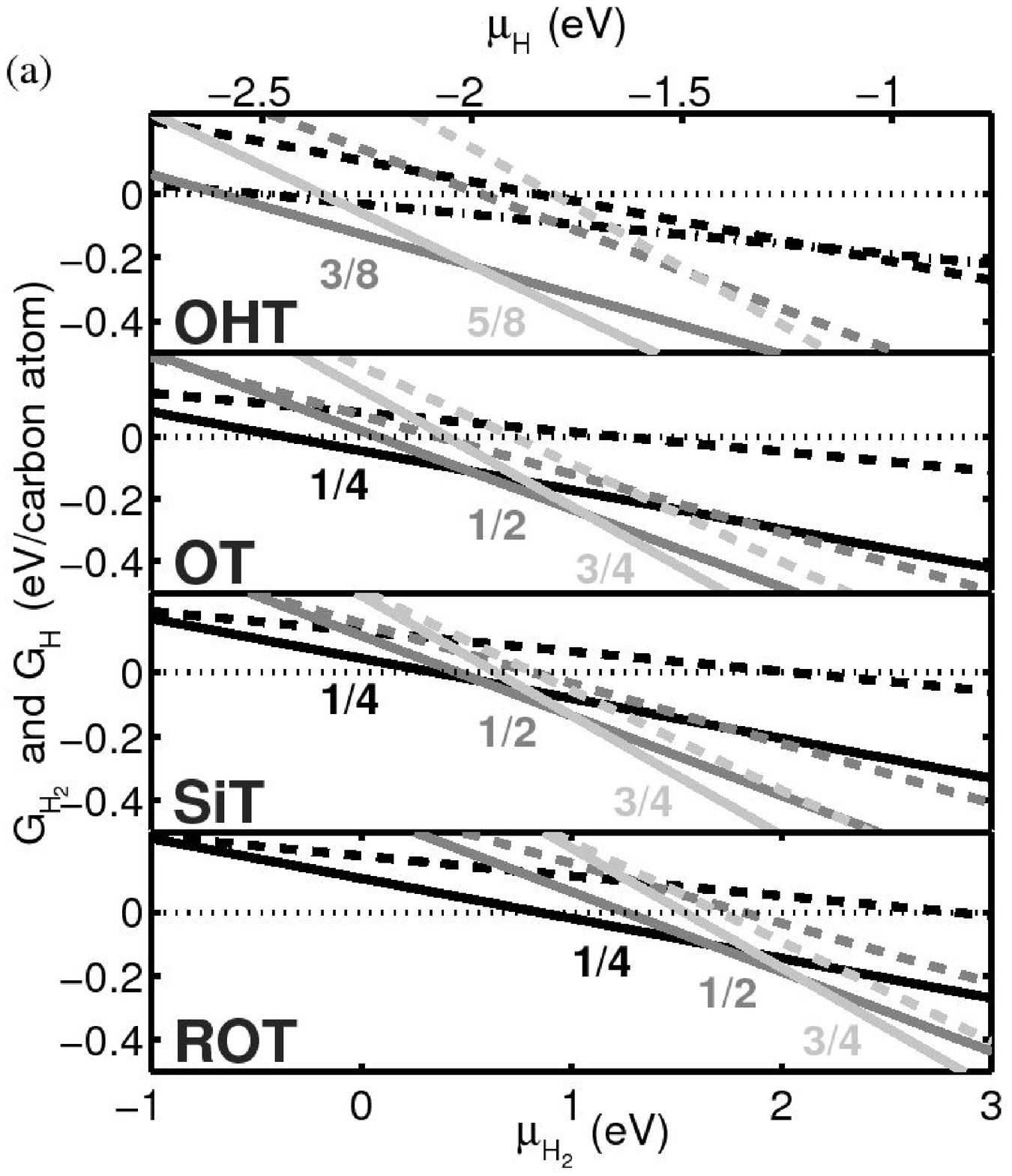} 
\includegraphics[width=0.47\textwidth]{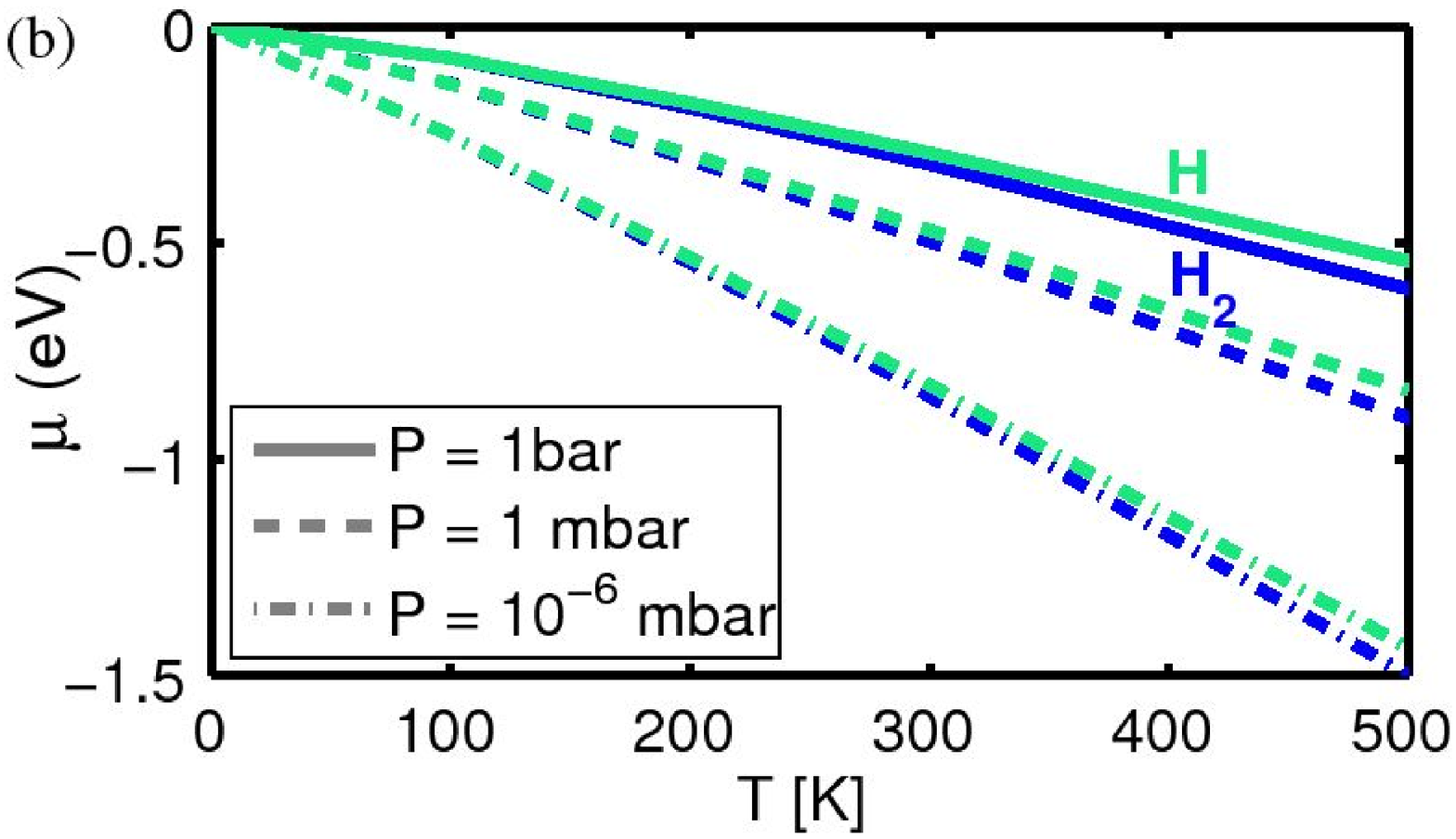}
\caption{\label{fig:deltaG} (a) The Gibbs energy of adsorption on the four surfaces as a function of the chemical potential of  molecular hydrogen (lower x-axis) and atomic hydrogen (upper x-axis). For positive $\Delta G_{ads}$, the unhydrogenated surface is favored. The preferred structures with the lowest $\Delta G_{ads}$ are marked with solid lines and the corresponding coverage as monolayers is shown.  (b) (Color online) The chemical potential of molecular hydrogen H$_2$ (blue/black) and atomic hydrogen (green/gray) in the gas phase as a function of temperature at three different partial pressures of the hydrogen species.}
\end{figure}

In contrast to the experiments, our calculations are performed in vacuum at 0~K. The experimental conditions, such as the hydrogen content of the environment, naturally affect the outcome of the hydrogenation. Using an \emph{ab initio}
thermodynamics approach,\cite{Wassmann, Rogal} we may address the relative stability of the different hydrogenated surfaces as a function of the chemical potential of the hydrogen species. The Gibbs energy of the hydrogen adsorption is calculated as
\begin{equation}
 \Delta G_{ads} = E_{B} + \frac{\rho_H}{2}\mu_{H_2}, 
\end{equation}
where $E_B$ is the binding energy per carbon atom of the considered surface, $\rho_H$ the surface concentration of adsorbed hydrogen (in the units of hydrogen atoms per graphene carbon) and $\mu_{H_2/H}=\mu_{H_2/H}^{\circ}(T)+k_BT\ln{\frac{P_{H_2/H}}{P^{\circ}}}$ is the chemical potential of hydrogen molecules or atomic hydrogen, where $P$ is the partial pressure of the hydrogen species. The standard chemical potential $\mu_{H_2}^{\circ}(T)$ can be obtained from thermodynamical reference data \cite{Chase} and is shown in Fig.~\ref{fig:deltaG}b.  We can also compare the hydrogenation through hydrogen molecules to atomic hydrogen by defining the binding energy and $\Delta G_{ads}$ with respect to atomic hydrogen in
gas phase.

$\Delta G_{ads}$ as a function of hydrogen chemical potential for each of the four surfaces is shown in Fig.~\ref{fig:deltaG}, considering both molecular (H$_2$) and atomic (H) hydrogen. The structure with lowest $\Delta G_{ads}$ is the most favorable. At high chemical potentials, the structures with a higher hydrogen content become
favored, even if their DFT binding energies are positive and the structures are thus unstable in vacuum at 0~K. Due to the logarithmic dependence of the chemical potential on the pressure, reaching positive values for the chemical requires high temperatures and pressures. Thus, the more densely hydrogenated structures are hard to reach in experiments using molecular hydrogen. The stability diagrams presented in Fig.~\ref{fig:deltaG} are, of course, idealistic and do not take into account thermal kinetics  but they could serve as a rough guideline to the experimental conditions required to achieve different hydrogen coverages. \\

\section{Conclusions}

In conclusion, we have studied graphene and its hydrogenated version, graphane, on the $\alpha$-quartz SiO$_2$
substrate, considering four different terminations of the (0001) surface. In our calculations, graphene does
not form bonds with the substrate atoms. Instead, the linear dispersion in the vicinity of the K and K' points of the first Brillouin zone is retained. 

We find that the surface termination has a profound effect on the hydrogen adsorption.  On the OHT and OT surfaces, stable coverages are found, with 3/8, 5/8 and 1/8~ML, and 1/4~ML hydrogen content, respectively.  The presence of hydrogen below the carbon plane facilitates the adsorption, stabilizing the hydrogenated structures and allowing higher surface coverages.  The higher coverages (3/8 and 5/8~ML) on the OHT surface are semiconductors with band gaps 3.5--3.9~eV, whereas on the OT surface at 1/4~ML coverage we observe a peculiar band structure with a single Dirac cone between the K and K' points. 

Our results indicate that in order to hydrogenate graphene deposited on SiO$_2$, it is beneficial to oxygenate the surface and passivate it with hydrogen before graphene deposition. As SiO$_2$ is amorphous, uneven hydrogen adsorption may be explained through surface domains of different termination. In addition, a small amount of hydrogen under the graphene layer is not likely to change the conducting properties of graphene. This means that partial dehydrogenation \cite{Sessi} of graphane on the hydrogenated surface is still an option and could be useful in nanoelectronics, for instance for forming nanoribbons in an insulating graphane matrix.\cite{Ijas,Byun}

We acknowledge the support from Aalto-Nokia collaboration and Academy of Finland through its Centers of Excellence Program (2006-2011). M. I. acknowledges the financial support from the Finnish Doctoral Programme in Computational Sciences FICS.

\bibliography{paper}

\begin{thebibliography}{10}
\bibitem{OHTdist}%
  \BibitemOpen
  \bibinfo {note} {For consistency, we measure the distance from the substrate
  from the uppermost substrate Si or O atoms also on the OHT surface, as the
  surface hydrogen may attach itself to the carbon layer when hydrogen is
  added.}%
  \bibAnnoteFile{Stop}{OHTdist}%
\end{thebibliography}

\end{document}